\documentclass[preprint,aps,prd,floatfix,nofootinbib,superscriptaddress]{revtex4}
\usepackage{graphicx}
\usepackage{amsmath}
\usepackage{amssymb}
\usepackage{bm}
\usepackage{epsfig}
\usepackage[normalem]{ulem}
\begin{document}

\preprint{TUM-EFT 109/18}
\title{\mbox{}\\[10pt] Inclusive decays of $\bm{\eta_c}$ and $\bm{\eta_b}$ at NNLO with large $\bm{n_f}$ resummation}
\author{Nora~Brambilla}
\affiliation{Physik-Department, Technische Universit\"at M\"unchen,
James-Franck-Stra{\ss}e 1, 85748 Garching, Germany}
\affiliation{Institute for Advanced Study, Technische Universit\"at M\"unchen,
Lichtenbergstra{\ss}e 2~a, 85748 Garching, Germany}
\affiliation{Excellence Cluster Universe, Technische Universit\"at M\"unchen,
Boltzmannstra{\ss}e 2, D-85748, Garching, Germany}
\author{Hee~Sok~Chung}
\affiliation{Physik-Department, Technische Universit\"at M\"unchen,
James-Franck-Stra{\ss}e 1, 85748 Garching, Germany}
\affiliation{Excellence Cluster Universe, Technische Universit\"at M\"unchen,
Boltzmannstra{\ss}e 2, D-85748, Garching, Germany}
\author{Javad~Komijani}
\altaffiliation[Present address:]{~School of Physics and Astronomy, University of Glasgow, Glasgow G12 8QQ, UK}
\affiliation{Physik-Department, Technische Universit\"at M\"unchen,
James-Franck-Stra{\ss}e 1, 85748 Garching, Germany}
\affiliation{Institute for Advanced Study, Technische Universit\"at M\"unchen,
Lichtenbergstra{\ss}e 2~a, 85748 Garching, Germany}
\noaffiliation
\date{\today}

\begin{abstract}
Based on the nonrelativistic QCD factorization theorem, 
we resum QCD corrections to the inclusive decay rate of $\eta_c$ and $\eta_b$
in the large-$n_f$ limit using bubble chain resummation. 
By employing dimensional regularization, we show explicitly the cancellation of
the infrared renormalon ambiguity in the factorization formula at leading order
in $v$ in the large-$n_f$ limit, where $v$ is the typical heavy quark velocity
inside the meson. 
We also make predictions of the ratio of the inclusive decay rate to the decay 
rate into two photons. By comparing our results with a fixed-order calculation 
we conclude that resummation of QCD corrections is crucial in making an
unambiguous prediction. We also find significant corrections beyond the 
large-$n_f$ limit for the decay of $\eta_c$, which may imply that 
QCD corrections need to be resummed beyond the large-$n_f$ limit to
make an accurate prediction of the decay rate. 
\end{abstract}

\maketitle

\section{Introduction}

Nowadays there is much experimental effort devoted to investigating the nature 
of heavy quarkonium states. Precision measurements of the properties of 
heavy quarkonia such as their masses, decays or transitions can be done at 
future and ongoing experiments like Belle~II, BESIII, and LHCb at CERN. 
A good understanding of the nature of  heavy quarkonium states is essential in exploring other
processes that involve heavy quarkonia such as their production in high energy collisions. 
Following the substantial success of the B factories~\cite{Bevan:2014iga}, 
the Belle~II experiment at KEK in Japan is going to collect 50 times more data
of what Belle obtained and will be able to investigate the properties of 
pseudoscalar bottomonium $\eta_b$ state and the corresponding charmonium $\eta_c$ state. 

One of the basic observables regarding heavy quarkonium is its inclusive decay rate. 
Theoretical predictions of the inclusive decay rate of $\eta_c$ and $\eta_b$ 
have long been pursued using nonrelativistic QCD (NRQCD), which is an effective theory providing
a factorization formalism that separates the perturbative short-distance contributions from the nonperturbative long-distance ones~\cite{Bodwin:1994jh}. 
The NRQCD factorization formula for the decay rate
of a heavy quarknoium is a sum over products of NRQCD long-distance matrix 
elements (LDMEs) and the corresponding short-distance coefficients (SDCs). 
The NRQCD power counting attributes to the LDMEs a specific scaling with $v$, where $v$ is the relative velocity between
the heavy quark and the heavy antiquark in the quarkonium.
The SDCs may be computed in perturbation theory. 
Therefore, the sum in the NRQCD factorization formula is an expansion in powers of $v$ and $\alpha_s$. 

There are two major obstacles in achieving theoretical predictions of the decay
rates of $\eta_c$ and $\eta_b$ with high accuracy. 
First, even though the expressions for the SDCs that contribute to the 
decay rates are available through order $v^7$ (relative order 
$v^4$)~\cite{Bodwin:2002hg, Brambilla:2008zg},%
\footnote{Here, we adopt the power counting rules in Ref.~\cite{Bodwin:1994jh}.}
the NRQCD LDMEs are generally not known very well beyond the one at leading order in $v$. Second, the perturbative
corrections to the SDCs, which are currently known to
next-to-next-to-leading order (NNLO) in the strong coupling constant 
$\alpha_s$, are uncomfortably large, hinting at a possible failure of the
convergence of the perturbation series. Especially, the nonconvergence of the
perturbation series may correspond to renormalon ambiguities that arise when computing diagrams in dimensional regularization 
and resumming the perturbation series by using the Borel transform. 

These difficulties can be partially overcome by considering $R$, the ratio of 
the inclusive decay rate to the electromagnetic decay rate to two photons. 
A considerable simplification occurs in $R$: not only it is independent of 
the leading-order LDME, but also the correction of relative order $\alpha_s^0 v^2$ 
cancels in $R$. The NRQCD factorization scale dependence, which arises in the 
corrections of relative order $\alpha_s^2$ and $\alpha_s v^2$, also cancel in
$R$. Finally, the renormalon ambiguities associated with the loop corrections
to the initial heavy quark-antiquark states also cancel in the ratio up to relative order $v^2$. 
Nevertheless, the renormalon ambiguities associated with the final-state 
gluons in the inclusive decay rate survive in $R$, so the perturbation 
series for $R$ still suffers from nonconvergence. 

In this paper, we consider the resummation of perturbative corrections to the
ratio $R$ that are associated with the chain of vacuum-polarization bubbles in the final-state gluons in the inclusive
decay rate of $\eta_Q$, where $Q = c$ or $b$. 
In Ref.~\cite{Bodwin:2001pt}, the resummation has been performed by imposing an 
infrared cutoff in the calculation of perturbative QCD corrections, which allows one to avoid
renormalon ambiguities that appear in the resummation if dimensional regularization is used instead.
In this work, we employ dimensional regularization to compute corrections in perturbative QCD in the limit of large number of active quark flavors $n_f$ 
and show explicitly in this limit the appearance of renormalon ambiguities in the perturbative QCD amplitude. 
We also show, by computing the perturbative corrections in NRQCD, that
perturbative NRQCD reproduces exactly the large $n_f$ leading renormalon ambiguity in
perturbative QCD, and therefore, the NRQCD factorization formula is free of this kind of renormalon ambiguities. 
We argue that, due to the limited knowledge on NRQCD LDMEs of higher orders in
$v$, using a hard cutoff instead of dimensional regularization to regularize 
the ultraviolet divergences in NRQCD leads to an expression for the 
decay rate of $\eta_Q$ that is more useful for phenomenological
applications. 
We also combine our resummed calculation with the perturbative 
calculation of $R$, which is currently known to NNLO in $\alpha_s$ and provide
updated numerical results for both $\eta_c$ and $\eta_b$. 

The paper is organized as follows. In Sec.~\ref{sec:resum} we consider the 
resummation of vacuum-polarization bubble chains in the $\eta_Q$ decay rate 
and obtain a resummed expression for $R$. We also combine the resummed result 
with the perturbative calculation of $R$. In Sec.~\ref{sec:numerical} we
present our numerical results for the ratio $R$ and compare them to
experimental data~\cite{Patrignani:2016xqp}. 
We conclude in Sec.~\ref{sec:summary}.

\section{Resummation of vacuum-polarization bubble chains in $R$}
\label{sec:resum}

In this section, we present the SDCs that contribute to the inclusive decay
rate of $\eta_Q$. To this end, we first shortly discuss the NRQCD 
factorization formula for the decay rate of $\eta_Q$. 
Then we use the factorization formula to
compute the decay rate of a perturbative $Q \bar Q$ state in perturbative QCD
and perturbative NRQCD. Finally, the SDCs are obtained by comparing the
expressions for the decay rate computed in QCD and NRQCD.

\subsection{NRQCD factorization for the decay rate of $\eta_Q$}
\label{sec:fact_formula}

The NRQCD factorization formula for the decay rate of $\eta_Q$, valid
through relative order $v^3$, reads~\cite{Bodwin:2002hg}
\begin{eqnarray}
\label{eq:fac-etaq}
\Gamma_{\eta_Q} &=& 
2 \, {\rm Im} \bigg[ \frac{f_1 ({}^1S_0)}{m^2} \bigg] 
\langle \eta_Q | {\cal O}_1({}^1S_0) | \eta_Q \rangle
+
2 \, {\rm Im} \bigg[ \frac{g_1 ({}^1S_0)}{m^4} \bigg] 
\langle \eta_Q | {\cal P}_1({}^1S_0) | \eta_Q \rangle
\nonumber \\ && + 
2 \, {\rm Im} \bigg[ \frac{f_8 ({}^3S_1)}{m^2} \bigg] 
\langle \eta_Q | {\cal O}_8({}^3S_1) | \eta_Q \rangle,
\end{eqnarray}
where $m$ is the pole mass of the heavy quark $Q$. 
The four-quark operators ${\cal O}_1({}^1S_0)$, 
${\cal P}_1({}^1S_0)$ and ${\cal O}_8({}^3S_1)$ are given by 
\begin{subequations}
\begin{eqnarray}
{\cal O}_1({}^1S_0) &=& \psi^\dag \chi \chi^\dag \psi, \\
{\cal P}_1({}^1S_0) &=& 
\frac{1}{2} [
\psi^\dag  \chi \chi^\dag (-\tfrac{i}{2} \overleftrightarrow{\bm{D}})^2\psi + 
\psi^\dag (-\tfrac{i}{2} \overleftrightarrow{\bm{D}})^2 \chi \chi^\dag \psi 
], \\
{\cal O}_8({}^3S_1) &=& \psi^\dag T^a \sigma^i \chi \chi^\dag T^a \sigma^i
\psi.
\end{eqnarray}
\end{subequations}
Here, $\psi$ and $\chi$ are the Pauli spinor field operators that annihilates
a heavy quark and creates a heavy antiquark, respectively.
The operator $\overleftrightarrow{\bm{D}}$ is the difference between the 
covariant derivative acting on the spinor to the right and on the spinor to the 
left, so that $\chi^\dag \overleftrightarrow{\bm{D}} \psi = 
\chi^\dag \bm{D} \psi - (\bm{D} \chi^\dag) \psi$.

The LDMEs 
$\langle \eta_Q | {\cal O}_1({}^1S_0) | \eta_Q \rangle$, 
$\langle \eta_Q | {\cal P}_1({}^1S_0) | \eta_Q \rangle$, and 
$\langle \eta_Q | {\cal O}_8({}^3S_1) | \eta_Q \rangle$ are 
nonperturbative quantities that correspond to the probabilities to find 
$Q \bar Q$ pairs in specific color and angular-momentum states in the $\eta_Q$ state. 
According to the power counting of Ref.~\cite{Bodwin:1994jh}, 
$\langle \eta_Q | {\cal O}_1({}^1S_0) | \eta_Q \rangle$, which scales
like $v^3$, is the LDME at leading order in $v$; 
$\langle \eta_Q | {\cal P}_1({}^1S_0) | \eta_Q \rangle$ is suppressed by $v^2$
compared to the leading-order LDME [the suppression comes from the two powers of derivatives in the operator ${\cal P}_1({}^1S_0)$] 
and scales like $v^5$; 
$\langle \eta_Q | {\cal O}_8({}^3S_1) | \eta_Q \rangle$, which scales
like $v^6$, is suppressed by $v^3$ 
compared to $\langle \eta_Q | {\cal O}_1({}^1S_0) | \eta_Q \rangle$.
The suppression of the LDME $\langle \eta_Q | {\cal O}_8({}^3S_1) | \eta_Q \rangle$ occurs 
because the operator ${\cal O}_8({}^3S_1)$ annihilates and creates $Q \bar Q$
in a color-octet state through a spin-flip process~\cite{Bodwin:1994jh}.

Power counting rules that are more conservative than those of Ref.~\cite{Bodwin:1994jh} 
have been suggested in Refs.~\cite{Brambilla:2001xy, Brambilla:2002nu}. 
In that power counting, which assumes $\Lambda_{\rm QCD}$ to be larger than $mv^2$, 
the LDME $\langle \eta_Q | {\cal P}_1({}^1S_0) | \eta_Q \rangle$ scales like before, i.e. like $m^2v^5$, 
while $\langle \eta_Q | {\cal O}_8({}^3S_1) | \eta_Q \rangle$ scales like $v^3\Lambda_{\rm QCD}^2/m^2$.
Moreover, there are two additional color-octet LDMEs that scale like $v^3 \Lambda_{\rm QCD}^2/m^2$ and $\Lambda_{\rm QCD}^2 v^3$ respectively, 
which are given by $\langle \eta_Q | {\cal O}_8({}^1S_0) | \eta_Q \rangle = \langle \eta_Q | \psi^\dag T^a \chi \chi^\dag T^a \psi | \eta_Q \rangle$
and 
$\langle \eta_Q | {\cal O}_8({}^1P_1) | \eta_Q \rangle
=\langle \eta_Q | \psi^\dag (-\frac{i}{2} \overleftrightarrow{\bm{D}}) T^a \chi 
\cdot
\chi^\dag (-\frac{i}{2} \overleftrightarrow{\bm{D}}) T^a \psi | \eta_Q \rangle$.
Therefore, if we adopt the power counting rules in Refs.~\cite{Brambilla:2001xy, Brambilla:2002nu}, 
Eq.~(\ref{eq:fac-etaq}) should include the above matrix elements to 
be valid through relative order $\Lambda_{\rm QCD}^2/m^2$. 
Considering, however, that in our numerical results, we will ignore the contribution from the 
color-octet LDME $\langle \eta_Q | {\cal O}_8({}^3S_1) | \eta_Q \rangle$ 
and account for its effect only in the uncertainties, and that the additional
color-octet LDMEs ignored in Eq.~(\ref{eq:fac-etaq}) 
do not affect the calculation of the renormalon ambiguities that we consider in
this paper, we conclude that we may consistently neglect also the additional color-octet LDMEs 
$\langle \eta_Q | {\cal O}_8({}^1S_0) | \eta_Q \rangle$ and $\langle \eta_Q | {\cal O}_8({}^1P_1) | \eta_Q \rangle$, 
whose effect is included in the uncertainties.
This is equivalent to assuming $mv \gg \Lambda_{\rm QCD}$ and restricting our calculation to a precision of relative order $v^2$.

The imaginary parts of the SDCs
$2 \, {\rm Im} [ f_1 ({}^1S_0)/m^2 ]$, 
$2 \, {\rm Im} [ g_1 ({}^1S_0)/m^4 ]$ and 
$2 \, {\rm Im} [ f_8 ({}^3S_1)/m^2 ]$ 
can be computed in perturbation theory. 
A general method to compute the SDCs is to consider 
Eq.~(\ref{eq:fac-etaq}) with the nonperturbative meson state replaced by the
perturbative $Q \bar Q$ state with definite color and angular momentum,
\begin{eqnarray}
\label{eq:fac-pert}
\Gamma_{Q \bar Q(n)} &=&  
2 \, {\rm Im} \bigg[ \frac{f_1 ({}^1S_0)}{m^2} \bigg] 
\langle Q \bar Q(n) | {\cal O}_1({}^1S_0) | Q \bar Q(n) \rangle
+
2 \, {\rm Im} \bigg[ \frac{g_1 ({}^1S_0)}{m^4} \bigg] 
\langle Q \bar Q(n) | {\cal P}_1({}^1S_0) | Q \bar Q(n) \rangle
\nonumber \\ && + 
2 \, {\rm Im} \bigg[ \frac{f_8 ({}^3S_1)}{m^2} \bigg] 
\langle Q \bar Q(n) | {\cal O}_8({}^3S_1) | Q \bar Q(n) \rangle . 
\end{eqnarray}
Here, $n$ denotes the color and angular momentum state of the $Q \bar Q$. 
We can compute the $\Gamma_{Q \bar Q}$ on the left-hand side in
perturbative QCD and compute the LDMEs on the right-hand side in perturbative
NRQCD for the $Q \bar Q$ states in various color, spin and
orbital angular momentum states. Then, the SDCs can be
determined by comparing the expressions on the left- and right-hand sides. 
In fixed-order perturbation theory, all SDCs
in Eq.~(\ref{eq:fac-etaq}) appear from order $\alpha_s^2$~\cite{Bodwin:1994jh}.
The SDC $2 \, {\rm Im} [ f_1 ({}^1S_0)/m^2 ]$ 
at leading order (LO) and next-to-leading order (NLO) in $\alpha_s$ 
has been computed in Refs.~\cite{Barbieri:1979be, Hagiwara:1980nv}, 
and the corrections at next-to-next-to-leading order (NNLO) in $\alpha_s$ have
been calculated recently in Ref.~\cite{Feng:2017hlu}.
The SDC $2 \, {\rm Im} [ g_1 ({}^1S_0)/m^4 ]$ at LO in $\alpha_s$ has been computed in
Ref.~\cite{Keung:1982jb}, and the corrections at NLO in $\alpha_s$ have 
been calculated in Ref.~\cite{Guo:2011tz}.
The SDC $2 \, {\rm Im} [ f_8 ({}^3S_1)/m^2 ]$ has been computed up to NLO
accuracy in $\alpha_s$ in Ref.~\cite{Petrelli:1997ge}. 

The analogous NRQCD factorization formula for the decay of $\eta_Q$ into two 
photons, valid through relative order $v^3$, reads%
\footnote{This expression is also valid, up to relative orders $v^2$ and $\Lambda_{\rm QCD}^2/m^2$,
in the more conservative power counting of Refs.~\cite{Brambilla:2001xy, Brambilla:2002nu}.}
\begin{equation}
\label{eq:fac-emdecay}
\Gamma_{\eta_Q\to \gamma \gamma} = 
2 \, {\rm Im} \bigg[ \frac{f_{\rm EM} ({}^1S_0)}{m^2} \bigg] 
\langle \eta_Q | {\cal O}_{\rm EM}({}^1S_0) | \eta_Q \rangle
+
2 \, {\rm Im} \bigg[ \frac{g_{\rm EM} ({}^1S_0)}{m^4} \bigg] 
\langle \eta_Q | {\cal P}_{\rm EM}({}^1S_0) | \eta_Q \rangle, 
\end{equation}
where the electromagnetic operators ${\cal O}_{\rm EM}({}^1S_0)$ and 
${\cal P}_{\rm EM}({}^1S_0)$ are given by 
\begin{subequations}
\begin{eqnarray}
{\cal O}_{\rm EM} ({}^1S_0) &=& \psi^\dag \chi |0\rangle \langle 0| 
\chi^\dag \psi, \\
{\cal P}_{\rm EM} ({}^1S_0) &=& 
\frac{1}{2} [
\psi^\dag  \chi |0\rangle \langle 0| 
\chi^\dag (-\tfrac{i}{2} \overleftrightarrow{\bm{D}})^2\psi + 
\psi^\dag (-\tfrac{i}{2} \overleftrightarrow{\bm{D}})^2 \chi 
|0\rangle \langle 0| \chi^\dag \psi 
]. 
\end{eqnarray}
\end{subequations}
Here, $|0\rangle$ is the QCD vacuum. 
The SDC $2 \, {\rm Im} [ f_{\rm EM} ({}^1S_0)/m^2 ]$ 
have been computed up to NNLO in $\alpha_s$ in fixed-order perturbation
theory~\cite{Barbieri:1979be, Czarnecki:2001zc, Feng:2015uha}, and
$2 \, {\rm Im} [ g_{\rm EM} ({}^1S_0)/m^4 ]$ is available up to 
NLO in $\alpha_s$~\cite{Keung:1982jb, Guo:2011tz, Jia:2011ah}.
The electromagnetic LDMEs
$\langle \eta_Q | {\cal O}_{\rm EM}({}^1S_0) | \eta_Q \rangle$ and 
$\langle \eta_Q | {\cal P}_{\rm EM}({}^1S_0) | \eta_Q \rangle$ can be
related to the color-singlet LDMEs 
$\langle \eta_Q | {\cal O}_1 ({}^1S_0) | \eta_Q \rangle$ and 
$\langle \eta_Q | {\cal P}_1 ({}^1S_0) | \eta_Q \rangle$ by using the
vacuum-saturation approximation, which holds up to corrections of relative
order $v^4$~\cite{Bodwin:1994jh},
\begin{subequations}
\begin{eqnarray}
\langle \eta_Q | {\cal O}_{\rm EM}({}^1S_0) | \eta_Q \rangle &=& 
\langle \eta_Q | {\cal O}_1 ({}^1S_0) | \eta_Q \rangle [ 1+O(v^4)],
\\
\langle \eta_Q | {\cal P}_{\rm EM}({}^1S_0) | \eta_Q \rangle &=& 
\langle \eta_Q | {\cal P}_1 ({}^1S_0) | \eta_Q \rangle [ 1+O(v^4)]. 
\end{eqnarray}
\end{subequations}
Putting Eqs.~(\ref{eq:fac-etaq}) and~(\ref{eq:fac-emdecay}) together, 
the NRQCD expression for the ratio $R$, valid up to relative order $v^3$, is 
\begin{eqnarray}
R &=& 
\frac{ {\rm Im} [ f_1 ({}^1S_0)/m^2 ]}
{ {\rm Im} [ f_{\rm EM} ({}^1S_0)/m^2 ] }
\bigg[ 
1+ 
\left( 
\frac{ {\rm Im} [ g_1 ({}^1S_0)/m^4 ]}
{ {\rm Im} [ f_1 ({}^1S_0)/m^2 ] }
-
\frac{ {\rm Im} [ g_{\rm EM} ({}^1S_0)/m^4 ]}
{ {\rm Im} [ f_{\rm EM} ({}^1S_0)/m^2 ] }
\right) 
\frac{\langle \eta_Q | {\cal P}_1 ({}^1S_0) | \eta_Q \rangle}
{\langle \eta_Q | {\cal O}_1 ({}^1S_0) | \eta_Q \rangle} 
\bigg] 
\nonumber \\ 
&& + \frac{ {\rm Im} [ f_8 ({}^3S_1)/m^2 ]}{ {\rm Im} [ f_{\rm EM} ({}^1S_0)/m^2 ] } 
\frac{\langle \eta_Q | {\cal O}_8 ({}^3S_1) | \eta_Q \rangle}{\langle \eta_Q | {\cal O}_1 ({}^1S_0) | \eta_Q \rangle} , 
\end{eqnarray}
where the second term in the square brackets corresponds to the correction at
relative order $v^2$, and the last term on the right-hand side gives the 
order-$v^3$
contribution. The order-$v^2$ correction to $R$ vanishes at LO in
$\alpha_s$; this is because the tree-level Feynman diagrams for 
$Q \bar Q \to gg$ and $Q \bar Q \to \gamma \gamma$ are same. 
The order-$\alpha_s v^2$ correction to $R$ can be obtained from the 
order-$\alpha_s v^2$ corrections to $\Gamma_{\eta_Q}$ and 
$\Gamma_{\eta_Q \to \gamma\gamma}$. The correction at order $\alpha_s v^2$ is
numerically small for both $\eta_c$ and $\eta_b$, and it is comparable to the
nominal size of the order-$v^3$ correction, which is often neglected (and included in the uncertainties) 
because the color-octet matrix element $\langle \eta_Q | {\cal O}_8 ({}^1S_0) | \eta_Q \rangle$ is not known.
We will follow this approach also here when providing numerical results (see Sec.~\ref{sec:numerical}), 
but we will keep the color-octet matrix element $\langle \eta_Q | {\cal O}_8 ({}^1S_0) | \eta_Q \rangle$ 
when discussing the renormalon cancellation in the rest of this section.

It is known from fixed-order calculations that the NLO and NNLO corrections to 
the SDCs ${\rm Im} [ f_1 ({}^1S_0)/m^2 ]$ and 
${\rm Im} [ f_{\rm EM} ({}^1S_0)/m^2 ]$ are large. Especially, there are
large corrections that are associated with the running of $\alpha_s$, where a
factor of $\alpha_s$ is accompanied by a factor of the QCD beta function. 
One way to resum (partially) such corrections is to consider chains of vacuum-polarization
bubbles, which reproduce fixed-order perturbation theory in the limit where the 
number of active quark flavors $n_f$ is large~\cite{Gross:1974jv,Lautrup:1977hs, thooft}.
In the ratio ${\rm Im} [ f_1 ({}^1S_0)/m^2 ]/{\rm Im} [ f_{\rm EM} ({}^1S_0)/m^2 ]$, 
the perturbative corrections at large $n_f$ that arise from initial-state
virtual gluons cancel~\cite{Bodwin:2001pt}. 
Therefore, in $R$, it suffices to consider only the
perturbative corrections to ${\rm Im} [ f_1 ({}^1S_0)/m^2 ]$ 
at large $n_f$ that arise from the final-state gluons. 
In the next part, we resum the QCD corrections to 
$2\, {\rm Im} [ f_1 ({}^1S_0)/m^2 ]$ that are associated with the final-state gluons in the large $n_f$ limit. 

The series for QCD corrections corresponding to bubble-chain diagrams in general do not converge.
If one attempts to make use of the Borel transform to carry out the resummation 
of such series,
the nonconvergence manifests itself through singularities
in the Borel plane. The inverse Borel transform becomes
ill-defined when the singularities reside on the 
positive axis of the Borel plane.
This gives rise to the so-called renormalon ambiguity in the resummed series.
The origin of the problem is that loop integrals contain contributions from regions of small gluon momenta
where perturbation theory breaks down (for QCD)~\cite{Bodwin:1998mn}.
In the factorization formula [Eq.~(\ref{eq:fac-pert})], 
loop integrals are partitioned so that contributions from small loop momenta are
contained in the LDMEs. Therefore, the SDCs
are free of infrared renormalon ambiguities if all possible LDMEs
are included in the factorization formula.
In practice, since we truncate the factorization formula at some orders in $v$, 
exact cancellations of renormalon ambiguities in the calculation of
the SDCs through the matching will occur through the order in $v$ at which the factorization formula is 
valid, and there will be remaining
ambiguities that are suppressed by powers of $v$. We shall demonstrate the
cancellation of leading renormalon ambiguities in the explicit calculation of $2 \, {\rm Im} [ f_1 ({}^1S_0)/m^2 ]$. 

Following Ref.~\cite{Bodwin:2001pt}, we employ two 
methods to carry out the bubble-chain resummation.
One method is na\"ive non-Abelianization (NNA), where we consider corrections to
the gluon propagator from $n_f$ light quark loops, and we promote the light-quark
part of the one-loop QCD beta function to the full one-loop QCD beta function
$\beta_0 = (33/2-n_f)/(6 \pi)$~\cite{Beneke:1994qe}. 
That is, we make the following replacement in the gluon propagator:
\begin{equation}
\frac{1}{k^2+i \varepsilon} \to K(x),
\end{equation}
where $x \equiv k^2/(4 m^2)$, $K(x) = \sum_{n=0}^\infty K^{(n)} (x)$ and 
\begin{equation}
K^{(n)}(x) = 
\frac{(\alpha_s \beta_0)^n [d-\log (-x-i \varepsilon)]^n}{4 m^2 
(x+i \varepsilon)}. 
\end{equation}
Here, $d$ is given by 
\begin{equation}
d = \log \frac{\mu^2}{4 m^2} - C, 
\end{equation}
where, in the $\overline{\rm MS}$ renormalization scheme, $C = - 5/3$ and $\mu$
is the renormalization scale. The strong coupling constant $\alpha_s$ is
also computed in the $\overline{\rm MS}$ scheme at the scale $\mu$. 
Another method is the background-field gauge (BFG) method, where the corrections to the gluon propagator from the gluon
and the ghost loops that are gauge dependent are also taken into account~\cite{DeWitt:1967ub}. 
In the BFG method in the $R_\xi$ gauge, $d$ is given by 
\begin{equation}
d = \log \frac{\mu^2}{4 m^2} + 
\frac{1}{\beta_0 \pi} 
\bigg[ \frac{67}{12} - \frac{5}{18} n_f - \frac{3}{4} (\xi^2-1) - \frac{5}{3}-C
\bigg],
\end{equation}
where $\xi$ is the gauge-fixing parameter. The choice $\xi=1$ corresponds to the Feynman gauge. 
If we set $\xi^2 = 7/3$, we reproduce the NNA method. Hence, the NNA
expression for the gluon propagator may be interpreted as a special case of the BFG expression for $\xi^2 = 7/3$.
In order to examine the dependence on the gauge-fixing parameter $\xi$,
we employ both the NNA method, which is equivalent to the BFG method for 
$\xi^2 = 7/3$, and the BFG method in the Feynman gauge ($\xi=1$). 

In the bubble-chain resummation, the left-hand side of Eq.~(\ref{eq:fac-pert}) 
occurs from order $\alpha_s$ through the decay into a single bubble-chain 
gluon. In order to decay into a virtual gluon, the $Q \bar Q$
pair must be in a color-octet state. If we take the $Q \bar Q$ pair to be in
the color-octet spin-triplet state and take the relative momentum between the
$Q$ and the $\bar Q$ to vanish, the matrix element 
$\langle Q \bar Q_8({}^3S_1)|{\cal O}_8({}^3S_1)|Q \bar Q_8({}^3S_1) \rangle$ 
occurs from order $\alpha_s^0$, and the matrix elements 
$\langle Q \bar Q_8({}^3S_1)|{\cal O}_1({}^1S_0)|Q \bar Q_8({}^3S_1) \rangle$ 
and
$\langle Q \bar Q_8({}^3S_1)|{\cal P}_1({}^1S_0)|Q \bar Q_8({}^3S_1) \rangle$ 
vanish through order $\alpha_s^0$. 
Hence, the SDC $2\,{\rm Im} [f_8({}^3S_1)/m^2]$ occurs form order $\alpha_s$, while the SDCs
$2 \, {\rm Im} [f_1 ({}^1S_0)/m^2]$ and 
$2 \, {\rm Im} [g_1 ({}^1S_0)/m^4]$ vanish through order $\alpha_s$. 
The SDC $2 \, {\rm Im} [f_8({}^3S_1)/m^2]$ has been computed in bubble-chain 
resummation in Ref.~\cite{Bodwin:2001pt} as
\begin{equation}
\label{eq:Im-f_8}
2 \, {\rm Im} \bigg[ \frac{f_8 ({}^3S_1)}{m^2} \bigg] = -8 \pi \alpha_s 
{\rm Im} [K(1)]. 
\end{equation}
In order to compute the SDC $2 \, {\rm Im} [f_1 ({}^1S_0)/m^2]$, 
we consider the left-hand side of Eq.~(\ref{eq:fac-pert}) at order $\alpha_s^2$. 
At this order, $\Gamma_{Q \bar Q (n)}$ occurs 
through the decay into two bubble-chain gluons. If we take the $Q \bar Q$ pair
to be in the color-singlet spin-singlet $S$-wave state, the matrix element 
$\langle Q \bar Q_1({}^1S_0)|{\cal O}_1({}^1S_0)|Q \bar Q_1({}^1S_0) \rangle$
occurs from order $\alpha_s^0$, while the matrix element 
$\langle Q \bar Q_1({}^1S_0)|{\cal O}_8({}^3S_1)|Q \bar Q_1({}^1S_0) \rangle$
occurs from order $\alpha_s$. If we take the relative momentum $q$ 
between the $Q$
and the $\bar Q$ to be zero, the matrix element 
$\langle Q \bar Q_1({}^1S_0)|{\cal P}_1({}^1S_0)|Q \bar Q_1({}^1S_0) \rangle$
vanishes through order $\alpha_s^0$. 
Since the SDCs $2 \, {\rm Im} [f_1 ({}^1S_0)/m^2]$ and
$2 \, {\rm Im} [g_1 ({}^1S_0)/m^4]$ appear from order $\alpha_s^2$,
while the SDC $2 \, {\rm Im} [f_8({}^3S_1)/m^2]$ occurs from order $\alpha_s$,
the contribution from the LDME
$\langle Q \bar Q_1({}^1S_0)|{\cal P}_1({}^1S_0)|Q \bar Q_1({}^1S_0) \rangle$
to the right-hand side of Eq.~(\ref{eq:fac-pert})
vanishes at order $\alpha_s^2$ when $q=0$. This implies that if we take the
$Q \bar Q$ state to be in the color-singlet ${}^1S_0$ state with $q=0$,
the SDC $2 \, {\rm Im} [g_1
({}^1S_0)/m^4]$ does not appear from the right-hand side of
Eq.~(\ref{eq:fac-pert}) at order $\alpha_s^2$,
and the right-hand side of Eq.~(\ref{eq:fac-pert})
involves at order $\alpha_s^2$ the SDCs $2 \, {\rm Im} [f_1 ({}^1S_0)/m^2]$
and $2 \, {\rm Im} [f_8 ({}^3S_1)/m^2]$.
Then, we can determine the SDC $2 \, {\rm Im} [f_1 ({}^1S_0)/m^2]$ 
by comparing $\Gamma_{Q \bar Q_1 ({}^1S_0)}$ [left-hand side of
Eq.~(\ref{eq:fac-pert})] with the right-hand side of 
Eq.~(\ref{eq:fac-pert}), where the LDMEs 
$\langle Q \bar Q_1({}^1S_0)|{\cal O}_1({}^1S_0)|Q \bar Q_1({}^1S_0) \rangle$
and 
$\langle Q \bar Q_1({}^1S_0)|{\cal O}_8({}^3S_1)|Q \bar Q_1({}^1S_0) \rangle$
are computed in perturbation theory.
The SDC $2 \, {\rm Im} [f_1 ({}^1S_0)/m^2]$ has been computed in
bubble-chain resummation in Ref.~\cite{Bodwin:2001pt} by regulating the
infrared divergences using a hard infrared cutoff on the virtuality of the 
final-state gluons. While using such an infrared cutoff effectively removes 
renormalon ambiguities in $2 \, {\rm Im} [f_1 ({}^1S_0)/m^2]$ by excluding 
contributions from arbitrarily soft gluon momenta, the cancellation of the
renormalon ambiguities in the factorization formula becomes obscure.

The appearance of the renormalon ambiguity in $\Gamma_{Q \bar Q (n)}$ and the
cancellation of the ambiguity in the SDC $2 \, {\rm Im} [f_1 ({}^1S_0)/m^2]$
can be seen explicitly by computing the SDC in dimensional regularization. 
In this section, we compute the SDC $2 \, {\rm Im} [f_1
({}^1S_0)/m^2]$ using bubble-chain resummation, by considering
Eq.~(\ref{eq:fac-pert}) where the the $Q \bar Q$ is in the color-singlet 
spin-singlet $S$-wave state with vanishing relative momentum between the $Q$ 
and the $\bar Q$. We regulate the infrared divergence
using dimensional regularization. 


\subsection{Computation in perturbative QCD}
\label{sec:resum_QCD}

\begin{figure}
\epsfig{file=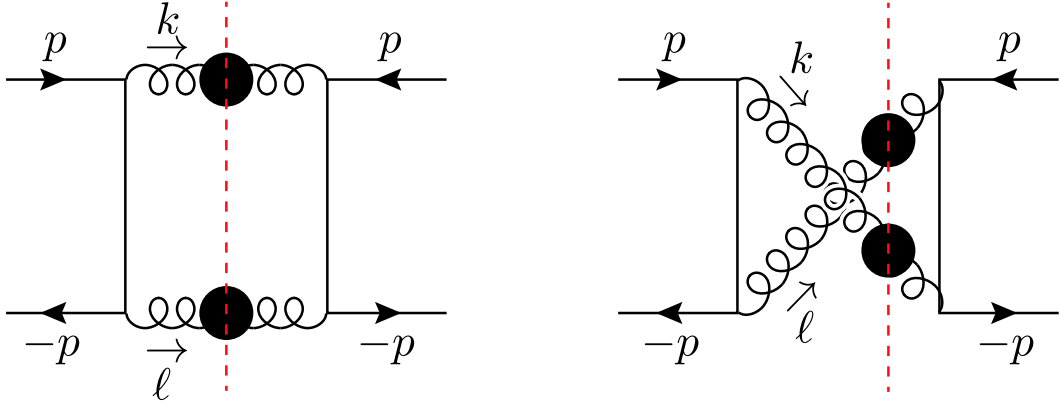,width=14cm}
\caption{
\label{fig:QCD_diagrams}
Feynman diagrams that contribute to $\Gamma_{Q \bar Q_1 ({}^1S_0)}$ at leading
order in $\alpha_s$ in perturbative QCD. 
Curly lines with filled circles represent bubble-chain gluons and the dashed
lines represent final-state cuts. 
}
\end{figure}

We compute the decay rate of a $Q \bar Q$ pair in the color-singlet ${}^1S_0$
state into two bubble-chain gluons. 
We use nonrelativistic normalization for the $Q \bar Q$ states. 
We set the momentum of the $Q$ and the $\bar Q$ to be $p$. 
To project the $Q \bar Q$ pair onto the color-singlet spin-singlet state, 
we replace the spinors by 
\begin{equation}
u(p) \bar v(p) \to 
\Pi_1 (p,p) \Lambda_1,  
\end{equation}
where $\Pi_1 (p,p)$ and $\Lambda_1$ are the spin-singlet and color-singlet
projectors, respectively~\cite{Kuhn:1979bb, Guberina:1980dc},
\begin{subequations}
\begin{eqnarray}
\Pi_1 (p,p) &=& - \frac{1}{2 \sqrt{2} m} (p\!\!\!/+m) \gamma_5, \\
\Lambda_1 &=& \frac{\bf 1}{\sqrt{N_c}}. 
\end{eqnarray}
\end{subequations}
Here, ${\bf 1}$ is the $SU(N_c)$ unit matrix. 
A straightforward calculation of the diagrams in Fig.~\ref{fig:QCD_diagrams} 
gives 
\begin{eqnarray}
\Gamma_{Q \bar Q_1 ({}^1S_0)} &=& 
\frac{1}{2}
\int \frac{d^4k}{(2 \pi)^4} 
\theta (k_0)
\int \frac{d^4\ell}{(2 \pi)^4} 
\theta (\ell_0)
(2 \pi)^4 \delta^{(4)} (2 p-k-\ell)  
2\, {\rm Im} [K (x)] \,
2\, {\rm Im} [K (y)] 
\nonumber \\ && \times 
\bigg|
{\rm tr} \bigg\{ \bigg[ 
(-i g \gamma^\nu T^b) 
\frac{i}{p \!\!\!/-k\!\!\!/-m+i \varepsilon} (-i g \gamma^\mu T^a) 
\nonumber \\ && \hspace{10ex} + 
(-i g \gamma^\mu T^a) 
\frac{i}{- p \!\!\!/+k\!\!\!/-m+i \varepsilon} (-i g \gamma^\nu T^b) 
\bigg]
\Pi_1 (p,p) \Lambda_1  
\bigg\} \bigg|^2, 
\end{eqnarray}
where $k$ and $\ell$ are the momenta of the final-state gluons, 
$g = \sqrt{4 \pi \alpha_s}$, 
$x \equiv k^2/(4 m^2)$, $y \equiv \ell^2/(4 m^2)$, 
and the trace is over the color and gamma matrices. 
Even though we employ dimensional regularization, it suffices to work in four
dimensions because in the current calculation, we encounter no 
divergences that require regularization. Then, 
\begin{eqnarray}
\Gamma_{Q \bar Q_1 ({}^1S_0)} 
&=&
\frac{C_F}{2} 
\frac{g^4}{2} 
\int \frac{d^4k}{(2 \pi)^4}
\theta (k_0)
\int \frac{d^4\ell}{(2 \pi)^4}
\theta (\ell_0)
(2 \pi)^4 \delta^{(4)} (2 p-k-\ell)
2\, {\rm Im} [K (x)] \, 
2\, {\rm Im} [K (y)] 
\nonumber \\ && \times 
\frac{16}{m^2} 
\frac{(k \cdot p)^2-m^2 k^2}{(k^2 -2 k \cdot p+i \varepsilon)^2}. 
\end{eqnarray}
We change the integration variables $k^0$, $\ell^0$ to $x$ and $y$,
so that from $x = (k_0^2-\bm{k}^2)/(4 m^2)$ and 
$y = (\ell_0^2-\bm{\ell}^2)/(4 m^2)$,
we obtain 
$d k_0 = 2 m^2 d x/k_0$ and 
$d \ell_0 = 2 m^2 d y/\ell_0$. Using the three-momentum components of 
the delta function to eliminate the integral over $\bm{\ell}$,
we replace $\bm{\ell}$ with $\bm{k}$, and we obtain
\begin{eqnarray}
\Gamma_{Q \bar Q_1 ({}^1S_0)} 
\label{eq:gam-qcd_0}
&=&
m^4
\frac{C_F}{2} 
\frac{g^4}{2} 
\int_0^1 \frac{d x}{\pi}
\int_0^1 \frac{dy }{\pi}
\int \frac{d^3\bm{k}}{(2 \pi)^3}
2 \pi \delta (2 m-k_0-\ell_0)
\frac{\theta (\ell_0)}{\ell_0}
\frac{\theta (k_0)}{k_0}
\nonumber \\ && \times 
2\, {\rm Im} [K (x)] \, 
2\, {\rm Im} [K (y)] 
\frac{16}{m^2} 
\frac{(k \cdot p)^2-m^2 k^2}{(k^2 -2 k \cdot p+i \varepsilon)^2}. 
\end{eqnarray}
Here, 
$k_0 = \sqrt{4 m^2 x + \bm{k}^2}$ and $\ell_0 = \sqrt{4 m^2 y + \bm{k}^2}$. 
The lower limits of the integrals over $x$ and $y$ are set by the fact that the
imaginary parts of $K(x)$ and $K(y)$ vanish for negative values of $x$ and $y$, 
respectively.
The upper limits of the integrals over $x$ and $y$ are set by the fact that the 
maximum invariant mass of a final-state particle is equal to the invariant mass
of the $Q \bar Q$ in the initial state.
Since we have chosen a root of the square root function such that $k_0 > 0$ and $\ell_0>0$,
we can drop $\theta(k_0)$ and $\theta(\ell_0)$. 
The remaining delta function in Eq.~(\ref{eq:gam-qcd_0})
constrains $\bm{k}^2$ to be $m^2 [1-2 (x+y)+(x-y)^2]$ and
$1-\sqrt{x}-\sqrt{y} \ge 0$;
we then obtain
\begin{eqnarray}
\label{eq:gam-qcd}
\Gamma_{Q \bar Q_1 ({}^1S_0)} &=& 
\frac{2 \pi C_F \alpha_s^2}{m^2} 
\sum_{n_1=0}^\infty 
\sum_{n_2=0}^\infty 
\int_0^1 \frac{dx}{2 \pi} 
\int_0^1 \frac{dy}{2 \pi} 
\, 2 \, {\rm Im} [4 m^2 K^{(n_1)} (x) ] 
\, 2 \, {\rm Im} [4 m^2 K^{(n_2)} (x) ] 
\nonumber \\ && \times 
f(x,y) \theta (1-\sqrt{x}-\sqrt{y}),
\end{eqnarray}
where 
\begin{equation}
\label{eq:f_x-y:def}
f(x,y) = \frac{[1-2 (x+y)+(x-y)^2]^{3/2}}{(1-x-y)^2}. 
\end{equation}
The sum over $n_1$ and $n_2$ corresponds to insertions of
$n_1$ and $n_2$ vacuum polarization bubbles to the two final-state gluon lines.
If we use the relation
\begin{equation}
\sum_{n=0}^\infty 
\int_0^1 \frac{dx}{2 \pi} 
\, {\rm Im} [4 m^2 K^{(n)} (x) ] 
F(x,y) 
= 
\sum_{n=0}^\infty (\alpha_s \beta_0)^n 
\left( \frac{d}{dt} \right)^n e^{td} \int_0^1 \frac{dx}{2 \pi} 
\, {\rm Im} \bigg[ \frac{x^{-t}
e^{i \pi t}}{x+i \varepsilon} \bigg] F(x,y) \bigg|_{t=0}, 
\end{equation}
which is valid for a generic function $F(x,y)$, 
we obtain
\begin{equation}
\label{eq:GammaQCD1}
\Gamma_{Q \bar Q_1 ({}^1S_0)} = 
\frac{2 \pi C_F \alpha_s^2}{m^2} 
\sum_{n_1=0}^\infty 
\sum_{n_2=0}^\infty 
(\alpha_s \beta_0)^{n_1+n_2} 
\left( \frac{d}{dt} \right)^{n_1}
\left( \frac{d}{d \tau} \right)^{n_2}
e^{d (t+\tau)} T(t,\tau) \bigg|_{t=\tau=0}, 
\end{equation}
where 
\begin{eqnarray}
\label{eq:T_t-tau:def}
T(t, \tau) = \frac{1}{\pi^2} \int_0^1 dx \int_0^1 dy \, 
{\rm Im} \bigg[ \frac{x^{-t} e^{i \pi t}}{x+i \varepsilon} \bigg]
{\rm Im} \bigg[ \frac{y^{-\tau} e^{i \pi \tau}}{y+i \varepsilon} \bigg]
f(x,y) \theta(1-\sqrt{x}-\sqrt{y}). 
\end{eqnarray}
We compute $T(t,\tau)$ in the Appendix. 
The summation in Eq.~(\ref{eq:GammaQCD1}) can be rewritten in integral
form by using the Borel summation formula: the Borel sum of $\sum_{n=0}^\infty
a_n x^n$ is given by the integral
\begin{equation}
\int_0^\infty \frac{dt}{x} e^{-t/x} \phi(t), 
\end{equation}
where $\phi(x) = \sum_{n=0}^\infty a_n x^n/n!$.
Using this formula, we rewrite Eq.~(\ref{eq:GammaQCD1}) as
\begin{equation}
\label{eq:GammaQCD2}
\Gamma_{Q \bar Q_1 ({}^1S_0)} = 
\frac{2 \pi C_F \alpha_s^2}{m^2} 
\frac{1}{(\alpha_s \beta_0)^2} 
\int_0^\infty dt \, \int_0^\infty d \tau \, e^{-w (t+\tau)} T(t, \tau), 
\end{equation}
where 
\begin{equation}
\label{eq:w:def}
w = \frac{1}{\alpha_s \beta_0} -d .
\end{equation}
Note that $w$ depends on the scale $\mu$
through dependence on $\alpha_s(\mu)$ and $d$. This dependence, however,
cancels at the level of one-loop running of $\alpha_s(\mu)$.
The prefactor $2 \pi C_F \alpha_s^2/m^2$ corresponds to 
$\Gamma_{Q \bar Q_1 ({}^1S_0)}$ at leading order in $\alpha_s$ in 
fixed-order perturbation theory. 

The function $T(t, \tau)$ is regular for $0\leq t<1$ and $0\leq \tau<1$. For $t\ge 1$ and
$\tau \ge 1$, 
there are singularities in $T(t,\tau)$ that make the value of the integral in 
Eq.~(\ref{eq:GammaQCD2}) ambiguous. The singularities in $T(t,\tau)$ that occur 
for the smallest values of $t$ or $\tau$ are at $t=1$ or $\tau=1$,
\begin{subequations}
\label{eq:T-singular}
\begin{eqnarray}
\lim_{t \to 1} (1-t) T(t, \tau) &=& - \frac{3}{\pi} \sin (\pi \tau), 
\\
\lim_{\tau \to 1} (1-\tau) T(t, \tau) &=& - \frac{3}{\pi} \sin (\pi t); 
\end{eqnarray}
\end{subequations}
see Eqs.~(\ref{eq:T-residue-at-tau=1}) and (\ref{eq:T-residue-at-t=1}).
These singularities give the leading renormalon ambiguities in 
$\Gamma_{Q \bar Q_1 ({}^1S_0)}$. 

One way to estimate the size of the leading renormalon ambiguity is to inspect
the difference between the results for the integral over $t$ and $\tau$ when
the integration contour is above the renormalon singularity and below the
renormalon singularity~\cite{Bodwin:1998mn}. 
From the residue theorem, the estimated ambiguity 
in $\Gamma_{Q \bar Q_1 ({}^1S_0)}$ that
arises from the leading renormalon singularity in $T(t, \tau)$ is 
\begin{eqnarray}
\label{eq:leading_ambiguity}
\delta \Gamma_{Q \bar Q_1 ({}^1S_0)} &\sim&
\bigg| 
2 \times 2 \pi i \times 
\frac{2 \pi C_F \alpha_s^2}{m^2} 
\frac{1}{(\alpha_s \beta_0)^2} 
\int_0^\infty d t\, e^{-w (1+t)} 
\frac{3}{\pi} \sin (\pi t) 
\bigg|
\nonumber \\
&=& 
\frac{2 \pi C_F \alpha_s^2}{m^2} 
\frac{12 \pi}{1-2 \alpha_s \beta_0 d + (\alpha_s \beta_0)^2 (\pi^2 + d^2)}
e^{-w}
.
\end{eqnarray}
For the case of $\eta_c$, the numerically estimated size of the leading renormalon 
ambiguity is of relative order one compared to 
$\Gamma_{Q \bar Q_1 ({}^1S_0)}$
at leading order in $\alpha_s$ in fixed-order perturbation theory.
This implies that for $\eta_c$, the value of the perturbation series 
$\Gamma_{Q \bar Q_1({}^1S_0)}$ has an ambiguity of order one.
Even for the case of $\eta_b$, the estimated ambiguity can be of relative 
order
$10^{-1}$, which is comparable to the nominal size of the order-$v^2$
corrections to the decay rate. 
Therefore, in order to make an accurate theoretical prediction of the $\eta_Q$ 
decay rate, it is crucial to have a factorization formula where such 
ambiguities are absent.

The renormalon ambiguities that arise from the singularities in $T(t,\tau)$
are located at $t = t_0$ or $\tau = t_0$ with $t_0=1$ being
the smallest and involves a factor 
$e^{-t_0 w}$. If we consider only the one-loop running of $\alpha_s$, this
factor can be written as 
\begin{equation}
e^{-t_0 w} \approx e^{t_0 d} \left( \frac{\Lambda_{\rm QCD}^2}{\mu^2}
\right)^{t_0}.
\end{equation}
Therefore, renormalon ambiguities that arise from the singularities in 
$T(t,\tau)$ that are located at larger values of $t$ or $\tau$ are suppressed 
by powers of $\Lambda_{\rm QCD}/\mu$. 
We can estimate the renormalon ambiguity from the first subleading
singularities in $T(t,\tau)$ which are located at $t=3/2$ or $\tau=3/2$, 
\begin{subequations}
\begin{eqnarray}
\lim_{t \to 3/2} (3/2-t) T(t, \tau) &=& \frac{2 (1-2 \tau)}{\pi} \sin (\pi \tau), 
\\
\lim_{\tau \to 3/2} (3/2-\tau) T(t, \tau) &=& \frac{2 (1-2 t)}{\pi} \sin (\pi t). 
\end{eqnarray}
\end{subequations}
The estimated renormalon ambiguity in $\Gamma_{Q \bar Q_1 ({}^1S_0)}$ 
from the first subleading singularities is 
\begin{eqnarray}
\label{eq:ambiguity_subleading}
\delta \Gamma_{Q \bar Q_1 ({}^1S_0)} &\sim&
\bigg| 
2 \times 2 \pi i \times 
\frac{2 \pi C_F \alpha_s^2}{m^2} 
\frac{1}{(\alpha_s \beta_0)^2} 
\int_0^\infty d t\, e^{-w (3/2+t)} 
\frac{2 (1-2 t)}{\pi} \sin (\pi t) 
\bigg|
\nonumber \\
&=& 
\frac{2 \pi C_F \alpha_s^2}{m^2} 
\frac{8 \pi [1-2 \alpha_s \beta_0 (2+d) + (\alpha_s \beta_0)^2 (\pi^2 +d^2 +4
d) ]}{ [ 1-2 \alpha_s \beta_0 d +(\alpha_s \beta_0)^2 (\pi^2 + d^2)]^2} 
e^{-\tfrac{3}{2}w}. 
\end{eqnarray}
This ambiguity is of relative order $10^{-1}$ for $\eta_c$ 
and is of relative order $10^{-3}$ 
for $\eta_b$. For $\eta_c$, the ambiguity is comparable to the nominal size of
the order-$v^4$ correction to the decay rate, and for $\eta_b$, the ambiguity
is smaller than the nominal size of the order-$v^4$ correction. 
Hence, the ambiguity from the subleading renormalon singularities in $T(t,
\tau)$ can be neglected at the current level of accuracy.

\subsection{Computation in perturbative NRQCD}
\label{sec:resum_NRQCD}

\begin{figure}
\epsfig{file=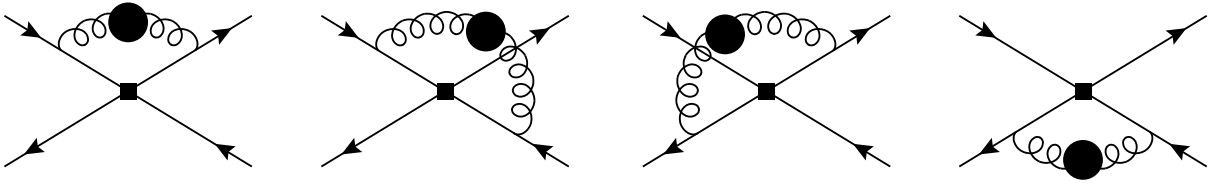,width=16cm}
\caption{
\label{fig:NRQCD_diagrams}
Feynman diagrams that contribute to 
$\langle {\cal O}_8({}^3S_1) \rangle_{Q \bar Q_1 ({}^1S_0)}$ 
at leading order in $\alpha_s$ in perturbative NRQCD. 
Curly lines with filled circles represent bubble-chain gluons and the filled squares represent the 
operator ${\cal O}_8({}^3S_1)$. 
}
\end{figure}

The renormalon ambiguities in the perturbation series of 
$\Gamma_{Q \bar Q_1 ({}^1S_0)}$ originate from integrations near zero loop 
momentum. In the NRQCD factorization formula [Eq.~(\ref{eq:fac-pert})],
the contributions from small momentum degrees of freedom are completely 
contained in the LDMEs. Hence, we expect the loop corrections to the NRQCD
LDMEs in the right-hand side of Eq.~(\ref{eq:fac-pert}), combined with the
SDCs, to reproduce the same renormalon ambiguities in $\Gamma_{Q \bar Q_1
({}^1S_0)}$. In this section, we compute the NRQCD LDMEs in 
Eq.~(\ref{eq:fac-pert}). 

Because we set the relative momentum between the $Q$ and the $\bar Q$ to be
zero, on the right-hand side of Eq.~(\ref{eq:fac-pert}), 
only the matrix element 
$\langle Q \bar Q_1 ({}^1S_0) | {\cal O}_1({}^1S_0) | Q \bar Q_1 ({}^1S_0)
\rangle$ appears at order $\alpha_s^0$.
At order $\alpha_s$, the matrix element
$\langle Q \bar Q_1({}^1S_0) | {\cal O}_8({}^3S_1) | Q \bar Q_1({}^1S_0) \rangle$ 
appears too. 
Since we only consider the NRQCD operators of the lowest mass
dimensions, the contributions from the right-hand side of
Eq.~(\ref{eq:fac-pert}) will only reproduce the leading renormalon ambiguity in
Eq.~(\ref{eq:GammaQCD2}).

At leading order in $\alpha_s$, the color-singlet LDME is given by 
\begin{equation}
\langle Q \bar Q_1 ({}^1S_0) | {\cal O}_1({}^1S_0) | Q \bar Q_1 ({}^1S_0)
\rangle = 2 N_c. 
\end{equation}
The color-octet matrix element vanishes at order $\alpha_s^0$, 
but it receives contributions at order $\alpha_s$ 
from the insertion of the
$\bm{\sigma} \cdot \bm{B}$ vertices to the quark and antiquark lines. 
The corresponding Feynman diagrams are shown in Fig.~\ref{fig:NRQCD_diagrams}. 
The sum of the diagrams gives 
\begin{equation}
\langle 
{\cal O}_8({}^3S_1) \rangle_{Q \bar Q_1 ({}^1S_0)}
= I \times 
\langle 
{\cal O}_1({}^1S_0) \rangle_{Q \bar Q_1 ({}^1S_0)}
+O(\alpha_s^2), 
\end{equation}
where
\begin{equation}
I = 
4 g^2 \frac{C_F}{2 N_c} 
\int \frac{d^4 k}{(2 \pi)^4} 
i K(x)
\left( \frac{1}{-k_0-\frac{\bm{k}^2}{2 m}+i \varepsilon} \right)^2
\frac{\bm{k}^2}{2 m^2} . 
\end{equation}
Here, we use the following shorthand notation 
$ \langle Q \bar Q_1 ({}^1S_0) | {\cal O}_1({}^1S_0) | Q \bar Q_1 ({}^1S_0) 
\rangle = \langle {\cal O}_1({}^1S_0) \rangle_{Q \bar Q_1 ({}^1S_0)}$ 
and
$ \langle Q \bar Q_1 ({}^1S_0) | {\cal O}_8({}^3S_1) | Q \bar Q_1 ({}^1S_0) 
\rangle = \langle {\cal O}_8({}^3S_1) \rangle_{Q \bar Q_1 ({}^1S_0)}$. 
If we rewrite $K(x)$ as 
\begin{equation}
i K^{(n)}(x) = 
-i (\alpha_s \beta_0)^n \left( \frac{d}{dt} \right)^n
\frac{(4 m^2)^t e^{td}}{(-k^2-i \varepsilon)^{1+t}} \bigg|_{t=0}, 
\end{equation}
we can deform the the contour for the integration over $k_0$ so that 
\begin{eqnarray}
I &=& 
 8 g^2 \frac{C_F}{2 N_c} 
\sum_{n=0}^\infty 
(\alpha_s \beta_0)^n \left( \frac{d}{dt} \right)^n
\frac{\pi
(4 m^2)^t e^{td} 
}{\Gamma(-t) \Gamma(1+t)}
\nonumber\\ && \times 
\int \frac{d^3 \bm{k}}{(2 \pi)^3}\int_{|\bm{k}|}^\infty \frac{dk_0}{2 \pi} 
\frac{1}{(k_0^2-\bm{k}^2)^{1+t}} 
\left( \frac{1}{-k_0-\frac{\bm{k}^2}{2 m} +i \varepsilon} \right)^2 
\frac{\bm{k}^2}{2 m^2}\bigg|_{t=0}. 
\end{eqnarray}
We first integrate over $k_0$. The result is 
\begin{equation}
\label{eq:int-k0}
\int_{|\bm{k}|}^\infty \frac{dk_0}{2 \pi} 
\frac{1}{(k_0^2-\bm{k}^2)^{1+t}} 
\left( \frac{1}{-k_0-\frac{\bm{k}^2}{2 m}+i \varepsilon} \right)^2 
= \frac{1}{2 \pi |\bm{k}|^{3+2 t}} J(t;\tfrac{|\bm{k}|}{2 m}), 
\end{equation}
where
\begin{equation}
J(t;z) = \frac{z}{ (1-z^2)^{t+2}}
\Gamma(t+2) \Gamma(-t) 
+
\frac{\Gamma (-t) \Gamma \left(t+\frac{3}{2}\right) \,
}{\Gamma(\frac{1}{2})}
F(1,t+\tfrac{3}{2};\tfrac{1}{2};z^2)
.
\end{equation}
Here, $F(a,b;c;z)$ is the hypergeometric function. 
Because we 
are matching QCD with NRQCD, we expand in $1/m$ and keep only the
contribution at leading power in $1/m$~\cite{Bodwin:1994jh},
\begin{eqnarray}
I &=& 
8 g^2 \frac{C_F}{2 N_c} 
\sum_{n=0}^\infty 
(\alpha_s \beta_0)^n \left( \frac{d}{dt} \right)^n
(4 m^2)^{t-1} e^{td} 
\frac{\Gamma(t+\frac{3}{2})}{\Gamma(1+t) \Gamma(\frac{1}{2})}
\int \frac{d^3 \bm{k}}{(2 \pi)^3} \frac{1}{|\bm{k}|^{1+2t}} 
\bigg|_{t=0}
\nonumber \\ &=& 
32 \pi \alpha_s \frac{C_F}{2 N_c} 
\frac{1}{\alpha_s \beta_0} \int_0^\infty dt \, 
e^{-w t} 
(4 m^2)^{t-1} 
\frac{\Gamma(t+\frac{3}{2})}{\Gamma(1+t) \Gamma(\frac{1}{2})}
\int \frac{d^3 \bm{k}}{(2 \pi)^3} \frac{1}{|\bm{k}|^{1+2t}} . 
\end{eqnarray}
In dimensional regularization, the integral over $\bm{k}$ is scaleless, and
hence vanishes,
\begin{eqnarray}
\label{eq:kintegral_DR}
\int \frac{d^3 \bm{k}}{(2 \pi)^3}
\frac{1}{|\bm{k}|^{1+2t}} 
&=&
\int \frac{d^3 \bm{k}}{(2 \pi)^3}
\frac{1}{|\bm{k}|^{1+2t}} 
= 
\frac{1}{2 \pi^2} 
\int_0^1 d |\bm{k}| \, |\bm{k}|^{1-2 t}
+ 
\frac{1}{2 \pi^2} 
\int_1^\infty d |\bm{k}| \, |\bm{k}|^{1-2 t}
\nonumber \\
&=& 
\frac{1}{4 \pi^2} 
\left( \frac{1}{1-t} 
- \frac{1}{1-t} \right), 
\end{eqnarray}
where in the second equality we split the integral over $|\bm{k}|$ 
so that the first (second) integral corresponds to the region where $|\bm{k}|$
is small (large). 
The first (second) integral is finite only when $t < 1$ ($t > 1$). 
After integrating over $|\bm{k}|$, we use analytical continuation to extend the
region of $t$ to the whole complex plane. 
Since 
the first term in the parenthesis comes from the region where $|\bm{k}|$
is small, and the second term originates from the region where $|\bm{k}|$ is
large, the first and second terms in the parenthesis correspond 
to the IR and UV renormalon singularities of the LDME $\langle {\cal O}_8
({}^3S_1)\rangle_{Q \bar Q_1 {({}^1S_0)}}$, respectively.
Then, we can write the right-hand side of Eq.~(\ref{eq:fac-pert}) as 
\begin{eqnarray}
\label{eq:GammaNRQCD_DR}
&& 
2 \, {\rm Im} \bigg[
\frac{f_1 ({}^1S_0)}{m^2} \bigg] 
\langle {\cal O}_1({}^1S_0) \rangle_{Q \bar Q_1 ({}^1S_0)}
+ 2 {\rm Im} \bigg[
\frac{f_8 ({}^3S_1)}{m^2} \bigg] 
\langle {\cal O}_8({}^3S_1) \rangle_{Q \bar Q_1 ({}^1S_0)}
\nonumber \\ && 
=
2 N_c \times 
2 \, {\rm Im} \bigg[
\frac{f_1 ({}^1S_0)}{m^2} \bigg] 
+ 
2 N_c \times 
\frac{\pi C_F \alpha_s^2}{N_c m^2} 
\frac{1}{(\alpha_s \beta_0)^2} \int_0^\infty dt \int_0^\infty d \tau \, 
e^{-w (t+\tau)} T_8^{\rm DR} (t,\tau)  , 
\end{eqnarray}
where 
\begin{equation}
T_8^{\rm DR} (t, \tau) = 
-\frac{3}{\pi} \sin (\pi \tau)
\left[ \frac{1}{(1-t)_{\rm IR}} - \frac{1}{(1-t)_{\rm UV}} \right]  
-\frac{3}{\pi} \sin (\pi t)
\left[ \frac{1}{(1-\tau)_{\rm IR}} - \frac{1}{(1-\tau)_{\rm UV}} \right].
\end{equation}
The subscripts IR and UV denote the origins of the IR and UV renormalon
singularities, respectively. 
Note that to derive Eq.~(\ref{eq:GammaNRQCD_DR}) we rewrote 
Eq.~(\ref{eq:Im-f_8}) as
\begin{equation}
2 \, {\rm Im} \bigg[ \frac{f_8 ({}^3S_1)}{m^2} \bigg] = - \frac{2 g^2}{4 m^2
\alpha_s \beta_0}
\int_0^\infty d\tau\, e^{-w \tau} \sin(\pi \tau) ,
\label{eq:Im-f_8:2}
\end{equation}
and symmetrized in $t$ and $\tau$.
By comparing Eq.~(\ref{eq:GammaNRQCD_DR}) with Eq.~(\ref{eq:GammaQCD2}), 
we can see that the infrared renormalon singularities in 
$T_8^{\rm DR} (t, \tau)$ [terms proportional to $1/(1-t)_{\rm IR}$ and
$1/(1-\tau)_{\rm IR}$] reproduce the leading 
renormalon singularities in $T (t,\tau)$ at $t=1$ or $\tau=1$, 
and, therefore, Eq.~(\ref{eq:GammaNRQCD_DR}) reproduces the leading renormalon
ambiguity in Eq.~(\ref{eq:GammaQCD2}). Then, the SDC 
$2\,{\rm Im} [f_1({}^1S_0)/m^2]$, given by 
\begin{equation}
2 \, {\rm Im} \bigg[
\frac{f_1 ({}^1S_0)}{m^2} \bigg] 
=
\frac{\pi C_F \alpha_s^2}{N_c m^2} 
\frac{1}{(\alpha_s \beta_0)^2} \int_0^\infty dt \int_0^\infty d \tau \, 
e^{-w (t+\tau)} [T(t,\tau)- T_8^{\rm DR} (t,\tau)] , 
\end{equation}
is free of the leading infrared renormalon ambiguity. 
On the other hand, the UV renormalon singularities in $T_8^{\rm DR} (t,\tau)$ 
[terms proportional to $1/(1-t)_{\rm UV}$ and $1/(1-\tau)_{\rm UV}$] has
no counterpart in perturbative QCD [Eq.~(\ref{eq:GammaQCD2})]; therefore, the 
SDC $2\, {\rm Im} [f_1({}^1S_0)/m^2]$ has a UV renormalon ambiguity. 
Since the UV renormalon ambiguity is absent in Eq.~(\ref{eq:GammaQCD2}), 
the UV renormalon ambiguities in the SDC $2\, {\rm Im} [f_1({}^1S_0)/m^2]$ 
and the LDME $\langle {\cal O}_8 ({}^3S_1)\rangle_{Q \bar Q_1 {({}^1S_0)}}$ 
cancel in the factorization formula Eq.~(\ref{eq:fac-pert}).
Since the UV renormalon ambiguities are of ultraviolet origin, the
nonperturbative LDME $\langle \eta_Q | {\cal O}_8 ({}^3S_1) | \eta_Q \rangle$
has the same UV renormalon ambiguities as the perturbative LDME $\langle {\cal
O}_8 ({}^3S_1)\rangle_{Q \bar Q_1 {({}^1S_0)}}$, with the perturbative
$Q \bar Q$ states replaced by the nonperturbative meson state. 
Therefore, the ambiguity is absent in the factorization formula for the
inclusive decay rate of $\eta_Q$ [Eq.~(\ref{eq:fac-etaq})]. 

Even though the UV renormalon ambiguities cancel in the factorization formula,
it is still necessary to define the SDC $2\, {\rm Im} [f_1({}^1S_0)/m^2]$ 
and the LDME $\langle \eta_Q | {\cal O}_8 ({}^3S_1) | \eta_Q \rangle$ 
unambiguously in order to compute the inclusive decay rate. An unambiguously
defined LDME will lead to an expression for $2\, {\rm Im} [f_1({}^1S_0)/m^2]$ 
that is free of UV renormalon ambiguities; however, different definitions will 
lead to different expressions for the SDC $2\, {\rm Im} [f_1({}^1S_0)/m^2]$ and 
the LDME $\langle \eta_Q| {\cal O}_8 ({}^3S_1)|\eta_Q \rangle$. 
Especially, the differences between different definitions of the 
LDME can be of the size of the UV renormalon ambiguity, which can be estimated
from the UV renormalon singularity of $\langle {\cal O}_8 ({}^3S_1)
\rangle_{Q \bar Q ({}^1S_0)}$ as
\begin{equation}
\langle \eta_Q | {\cal O}_1 ({}^1S_0) | \eta_Q \rangle 
\times 
\left|
2 \times 2 \pi i 
\frac{6 \alpha_s}{\pi} 
\frac{C_F}{2 N_c} 
\frac{1}{\alpha_s \beta_0} 
e^{-w} 
\right|
= \frac{16}{3 \beta_0} e^{-w} 
\langle \eta_Q | {\cal O}_1 ({}^1S_0) | \eta_Q \rangle.
\end{equation}
We note that the renormalon ambiguity scales like 
$(\Lambda_{\rm QCD}/\mu)^2$, which is different from 
velocity-scaling rules of the LDMEs~\cite{Bodwin:1998mn}. 
Hence, there is a possibility that the 
renormalon ambiguity in the LDMEs can spoil the expansion in powers of $v$
in case the renormalon ambiguity of an LDME exceeds its nominal size. 

We can define NRQCD LDMEs so that the LDMEs are free of UV renormalon 
ambiguities and also respect the velocity-scaling rules by regulating the UV
divergences using a cutoff regulator. In perturbative calculations, it is most
convenient to apply a hard cutoff $\Lambda$ on the size of the spatial momentum 
of the gluon. The cutoff $\Lambda$ should be large enough so that it encompasses 
the relevant momentum regions in NRQCD, while $\Lambda < m$ so that the 
expansion in powers of $1/m$ is valid. 
Hence, it is customary to choose $\Lambda \sim mv$. 
While the NRQCD LDMEs in hard-cutoff regularization are free of 
renormalon ambiguities, they depend on the cutoff $\Lambda$. 


If we regularize the UV divergences in 
NRQCD with a hard cutoff $\Lambda$ for the spatial momentum of the gluon, the 
integral over $\bm{k}$ in Eq.~(\ref{eq:kintegral_DR}) now becomes 
\begin{equation}
\int^\Lambda \frac{d^3 \bm{k}}{(2 \pi)^3}  \frac{1}{\bm{k}^{1+2t}} 
= 
\frac{4 \pi}{(2 \pi)^3}
\int_0^\Lambda d |\bm{k}| \, |\bm{k}|^{1-2 t}
= 
\frac{1}{4 \pi^2} 
\frac{\Lambda^{2-2 t}}{1-t}, 
\end{equation}
which yields
\begin{equation}
I^{(\Lambda)}
=
\frac{8 \alpha_s}{\pi} \frac{C_F}{2 N_c} 
\frac{1}{\alpha_s \beta_0} \int_0^\infty dt\, e^{-wt} 
\frac{\Gamma(t+\frac{3}{2})}{\Gamma(1+t) \Gamma(\frac{1}{2})}
\left( \frac{\Lambda}{2 m} \right)^{2-2 t}
\frac{1}{1-t}. 
\end{equation}
Here, the superscript ${(\Lambda)}$ denotes that a UV cutoff was used. 
Then, the right-hand side of Eq.~(\ref{eq:fac-pert}) reads
\begin{eqnarray}
\label{eq:GammaNRQCD}
&& 
2 \, {\rm Im} \bigg[
\frac{f_1 ({}^1S_0)}{m^2} \bigg] 
\langle {\cal O}_1({}^1S_0) \rangle_{Q \bar Q_1 ({}^1S_0)}
+ 2 \, {\rm Im} \bigg[
\frac{f_8 ({}^3S_1)}{m^2} \bigg] 
\langle {\cal O}_8({}^3S_1) \rangle_{Q \bar Q_1 ({}^1S_0)}
\nonumber \\ && 
=
2 N_c \times 
2 \, {\rm Im} \bigg[
\frac{f_1 ({}^1S_0)}{m^2} \bigg] 
+ 
2 N_c \times 
\frac{\pi C_F \alpha_s^2}{N_c m^2} 
\frac{1}{(\alpha_s \beta_0)^2} \int_0^\infty dt \int_0^\infty d \tau \, 
e^{-w (t+\tau)} T_8^{(\Lambda)} (t,\tau)  , 
\end{eqnarray}
where
\begin{eqnarray}
T_8^{(\Lambda)} (t, \tau) &=& 
\frac{\sin (\pi \tau)}{\pi}
\frac{-4\Gamma(t+\frac{3}{2})}{\Gamma(1+t) \Gamma(\frac{1}{2})}
\left( \frac{\Lambda}{2 m} \right)^{2-2 t}
\frac{1}{1-t}
\nonumber \\ && 
+ 
\frac{\sin (\pi t)}{\pi}
\frac{-4\Gamma(\tau+\frac{3}{2})}{\Gamma(1+\tau) \Gamma(\frac{1}{2})}
\left( \frac{\Lambda}{2 m} \right)^{2-2 \tau}
\frac{1}{1-\tau}. 
\end{eqnarray}
As we have discussed in Sec.~\ref{sec:fact_formula}, 
the SDC $2\, {\rm Im} [g_1({}^1S_0)/m^4]$ does not appear in 
Eq.~(\ref{eq:GammaNRQCD}) because the contribution from the LDME 
$\langle Q \bar Q_1({}^1S_0)|{\cal P}_1({}^1S_0)|Q \bar Q_1({}^1S_0) \rangle$ 
to the right-hand side of Eq.~(\ref{eq:fac-pert}) vanishes at order
$\alpha_s^2$ if the $Q \bar Q$ is in the color-singlet ${}^1S_0$ state and the 
relative momentum between the $Q$ and the $\bar Q$ is zero.
The singularities of $T_8^{(\Lambda)} (t,\tau)$ at $t=1$ or $\tau=1$ are given
by 
\begin{subequations}
\label{eq:T8-singular}
\begin{eqnarray}
\lim_{t \to 1} (1-t) T_8^{(\Lambda)} (t, \tau) &=& 
- \frac{3}{\pi} \sin (\pi \tau), \\
\lim_{\tau \to 1} (1-\tau) T_8^{(\Lambda)} (t, \tau) &=& 
- \frac{3}{\pi} \sin (\pi t), 
\end{eqnarray}
\end{subequations}
which reproduce the leading renormalon singularities in $T(t,\tau)$ at $t=1$ or
$\tau=1$. It is clear that, from the expression for $T_8^{(\Lambda)} (t,\tau)$,
these are the only singularities at $t>0$ and $\tau>0$. Therefore, we define 
the NRQCD LDME with a UV cutoff $\Lambda$ to obtain unambiguous expressions for
the SDC $2\, {\rm Im} [f_1 ({}^1S_0)/m^2]$.

\subsection{Summary of results}
\label{sec:resum_summary}

Here we summarize our result for the SDC
$2\, {\rm Im} [f_1 ({}^1S_0)/m^2]$. 
The left-hand side of Eq.~(\ref{eq:fac-pert}), computed in perturbative QCD for
the $Q \bar Q$ in the color-singlet spin-singlet $S$-wave state, is given in
Eq.~(\ref{eq:GammaQCD2}). The right-hand side computed in perturbative NRQCD is
given in Eq.~(\ref{eq:GammaNRQCD_DR}), when dimensionally regularized, and in
Eq.~(\ref{eq:GammaNRQCD}), when a hard cutoff is employed.
We continue with the latter, 
which does not require any further subtraction
of UV renormalons in perturbative NRQCD.
Then, by comparing Eq.~(\ref{eq:GammaQCD2}) with Eq.~(\ref{eq:GammaNRQCD}),
we obtain
\begin{equation}
\label{eq:Imf1cutoff}
2\, {\rm Im} \bigg[
\frac{f_1 ({}^1S_0)}{m^2} \bigg]
=
\frac{\pi C_F \alpha_s^2}{N_c m^2} 
\frac{1}{(\alpha_s \beta_0)^2} \int_0^\infty dt \int_0^\infty d \tau \, 
e^{-w (t+\tau)} [ T (t,\tau) - T_8^{(\Lambda)} (t,\tau) ] . 
\end{equation}
Since the functions $T (t,\tau)$ and $T_8^{(\Lambda)} (t,\tau)$ have same
singularities at $t=1$ or $\tau=1$ [Eqs.~(\ref{eq:T-singular}, 
\ref{eq:T8-singular})], 
those singularities cancel in 
$T (t,\tau) - T_8^{(\Lambda)} (t,\tau)$. Therefore, the leading renormalon 
ambiguities in $\Gamma_{Q \bar Q_1({}^1S_0)}$ that originate from the 
singularities at $t=1$ or $\tau=1$ are absent
in the SDC $2\, {\rm Im} [f_1 ({}^1S_0)/m^2]$. 

Together with our results for $2\,{\rm Im} [f_1 ({}^1S_0)/m^2]$ 
and the perturbative expression for 
$2\,{\rm Im} [f_{\rm EM} ({}^1S_0)/m^2]$ at LO in
$\alpha_s$~\cite{Bodwin:1994jh}, 
we obtain the resummed expression for $R$ at leading order in $v$ 
including resummed QCD corrections in the large $n_f$ limit,
\begin{equation}
\label{eq:R-resum}
R^{\rm Resum} = 
R_0 [1+O(v^2)] 
\times 
\frac{1}{(\alpha_s \beta_0)^2} \int_0^\infty dt \int_0^\infty d \tau \, 
e^{-w (t+\tau)} [ T (t,\tau) - T_8^{(\Lambda)} (t,\tau) ], 
\end{equation}
where 
\begin{equation}
R_0 = 
\frac{C_F \alpha_s^2}{2 N_c \alpha^2 e_Q^4}. 
\end{equation}
Here $e_Q$ is the fractional electric charge of the heavy quark $Q$.
As previously discussed, the order-$v^2$ correction to $R$ vanishes at LO in
$\alpha_s$. 
We neglect the correction at order $\alpha_s v^2$ that 
was computed in fixed-order perturbation theory, because it was found
to be small numerically~\cite{Guo:2011tz, Jia:2011ah} and is of 
comparable size to the contribution of order $v^3$.
The order-$v^3$ contribution to $R$ can be written as 
\begin{equation}
\label{eq:R8}
R_8 
= -2 \, {\rm Im} [4 m^2 K(1) ] 
R_0 \frac{N_c}{\alpha_s C_F} \frac{
\langle \eta_Q| {\cal O}_8({}^3S_1) |\eta_Q \rangle }
{\langle \eta_Q| {\cal O}_1({}^1S_0) |\eta_Q \rangle }. 
\end{equation}
Since it is not known how to compute the color-octet LDME 
$\langle \eta_Q| {\cal O}_8({}^3S_1) |\eta_Q \rangle$ reliably,
we ignore $R_8$ and instead consider its effects in the uncertainties. 

We can combine our results for $R^{\rm Resum}$ with fixed-order calculations of
$R$, so that the corrections at NLO and NNLO in $\alpha_s$ are valid beyond the
large $n_f$ limit. 
By using the expressions for $\Gamma_{\eta_Q}$ and 
$\Gamma_{\eta_Q \to \gamma \gamma}$ valid to NNLO in $\alpha_s$, we
obtain~\cite{Czarnecki:2001zc, Feng:2015uha, Feng:2017hlu} 
\begin{eqnarray}
\label{eq:R-pert}
R^{\rm Pert} &=& 
R_0 \bigg\{ 
1+ \frac{\alpha_s}{\pi} \bigg[ 
2 \pi \left(\beta_0 - \frac{n_H}{6 \pi} \right) \log \frac{\mu^2}{4 m^2} 
+ \hat R^{(1)} 
\bigg] 
\nonumber\\ && \hspace{6ex}
+ \left(\frac{\alpha_s}{\pi} \right)^2 \bigg[
3 \pi^2 \left(\beta_0 - \frac{n_H}{6 \pi} \right)^2 \log^2 \frac{\mu^2}{4 m^2} 
+ \hat R^{(2)} 
\nonumber\\ && \hspace{16ex}
+ \bigg( 2 \pi^2 \beta_1 - \frac{19}{12} n_H 
 + 3 \pi \left( \beta_0 - \frac{n_H}{6 \pi} \right) 
\hat R^{(1)} \bigg) \log \frac{\mu^2}{4 m^2} 
\bigg] 
+ O(\alpha_s^3, \alpha_s v^2, v^3) \bigg\},
\nonumber \\
\end{eqnarray}
where $\hat R^{(1)} = \left( \frac{199}{18} - \frac{13 \pi^2}{24} \right)
C_A - \frac{8}{9} n_f - \frac{2 n_H}{3} \log 2$, 
$\beta_1 = \frac{1}{(4 \pi)^2} 
\left( \frac{34}{3} C_A^2- \frac{20}{3} C_A T_R n_f - 4 C_F T_R n_f \right)$, 
and $n_H$ is the number of heavy quark flavors. 
$\hat R^{(2)}$ is known as a function of $n_f$ for the case $n_H = 1$ only. 
$\hat R^{(2)} = 117.144$ for $n_f = 3$ and 
$\hat R^{(2)} = 86.421$ for $n_f = 4$. 
The large $n_f$ limit of $\hat R^{(2)}$ is given in Ref.~\cite{Feng:2017hlu} 
as $\lim_{n_f \to \infty} \hat R^{(2)}/n_f^2 = 0.37581(3)$. 
The full $n_f$ dependence of $\hat R^{(2)}$ can be obtained from 
Refs.~\cite{Feng:2015uha, Feng:2017hlu} as 
\begin{equation}
\hat R^{(2)} = 
191.3 - \hat R^{(2)}_{\rm lbl} - 25.07 n_f + 0.3758 n_f^2,
\end{equation}
where $\hat R^{(2)}_{\rm lbl} = 
0.7313 \times C_F\times \sum_f (e_f/e_Q)^2 
+ 0.6470 \times C_F n_H$ 
is the ``light by light'' contribution to the two-photon decay rate 
that occurs through $Q \bar Q \to gg \to \gamma \gamma$ via a light quark loop. 
Here, the sum is over $n_f$ light quark flavors, and 
$e_f$ is the fractional charge of a light quark of flavor $f$. 

When $n_H = 1$, the heavy quark $Q$ contributes to the renormalization scale
dependence of $\alpha_s$, which cancels the explicit renormalization scale
dependence of $R^{\rm Pert}$ from the logarithms of $\mu/m$. 
It is possible to decouple the heavy quark $Q$ from the running of $\alpha_s$ 
by using the 
decoupling relations between $\alpha_s$ for $n_f$ and $n_f+1$ active quark
flavors~\cite{Chetyrkin:2000yt}, so that the heavy quark $Q$ does not affect
the renormalization scale dependence of $R^{\rm Pert}$ for $\mu < m$. 
By using Eq.~(25) of Ref.~\cite{Chetyrkin:2000yt}, we decouple the heavy
quark, and then we obtain
\begin{eqnarray}
\label{eq:R-pert-decoupled}
R^{\rm Pert} &=& 
R_0 \bigg\{ 
1+ \frac{\alpha_s}{\pi} \bigg[ 
2 \pi \beta_0 \log \frac{\mu^2}{4 m^2} 
+ \hat R'^{(1)} 
\bigg] 
\nonumber\\ && \hspace{6ex}
+ \left(\frac{\alpha_s}{\pi} \right)^2 \bigg[
3 \pi^2 \beta_0 ^2 \log^2 \frac{\mu^2}{4 m^2} 
+ \hat R'^{(2)} 
\nonumber\\ && \hspace{16ex}
+ \bigg( 2 \pi^2 \beta_1 + 3 \pi \beta_0 
\hat R^{(1)} \bigg) \log \frac{\mu^2}{4 m^2} 
\bigg] 
+ O(\alpha_s^3, \alpha_s v^2, v^3) \bigg\},
\end{eqnarray}
where $\hat R'^{(1)} = \left( \frac{199}{18} - \frac{13 \pi^2}{24} \right)
C_A - \frac{8}{9} n_f$ and $\hat R'^{(2)} = \hat R^{(2)}
+ \frac{7}{12} + \frac{19}{6} \log 2 - \frac{1}{3} \log^2 2+ \log 2 \, 
\hat R'^{(1)}$.
We use this expression for $R^{\rm Pert}$ when $\mu < m$.

The expression for $R^{\rm Pert}$ in Eq.~(\ref{eq:R-pert}) is obtained 
from the electromagnetic and inclusive decay rates of $\eta_Q$ that were
calculated in the $\overline {\rm MS}$ renormalization scheme. In order to make
Eq.~(\ref{eq:R-pert}) compatible with the expression for $R^{\rm Resum}$ in 
Eq.~(\ref{eq:R-resum}), it is necessary to convert Eq.~(\ref{eq:R-pert}) to the
hard cutoff scheme. It is possible to perform a finite renormalization of the
NRQCD LDMEs from the $\overline {\rm MS}$ scheme to the cutoff scheme. 
At leading order in $v$, the finite renormalization only involves the 
color-singlet LDME, and the finite renormalization cancels trivially in the 
ratio $R$. Even if we include contributions from the order-$v^2$ LDME, 
which contributes to $\langle \eta_Q | {\cal O}_1 ({}^1S_0) |\eta_Q \rangle$ 
at order
$\alpha_s$, the expression in Eq.~(\ref{eq:R-pert}) remains unchanged if the
corrections of order $\alpha_s^2 v^2$ and of order $v^4$ are ignored. 

Because Eq.~(\ref{eq:R-pert}) is computed by using dimensional regularization, 
all power divergences are absent in Eq.~(\ref{eq:R-pert}). 
In the fixed-order perturbation theory calculation using the 
cutoff-regularization scheme, the color-singlet contribution to $R$ receives
power-divergent contribution from the one-loop correction to the color-octet 
LDME, which is of relative order $\alpha_s \Lambda^2/m^2$. If we set $\Lambda
\sim m v$, this contribution is of relative order $\alpha_s v^2$. 
For $\alpha_s \sim v$, this is the same size as the color-octet contribution to
$R$, whose relative size is of order $v^3$; for $\alpha_s \sim v^2$, the
power-divergent contribution is of relative order $v^4$. Therefore, such 
power-divergent contributions in $R^{\rm Pert}$ can be ignored at the current 
level of accuracy. 

In order to combine the perturbative expression [Eq.~(\ref{eq:R-pert})] with 
the resummed result [Eq.~(\ref{eq:R-resum})], we need to
subtract from $R^{\rm Resum}$ the contributions 
that are already included in the perturbative expression $R^{\rm Pert}$
in order to avoid double counting. Since, in the perturbative calculation, the
contribution from the color-octet matrix element is not included, we only need
to consider the contribution from $T(t, \tau)$. From the series expansion of 
$T(t, \tau)$ at $t=\tau=0$ we find 
\begin{equation}
\label{eq:R-subtract}
\delta R^{\rm Resum} = 
R_0 \bigg[
1 + 2 (1+d) \alpha_s \beta_0 
+ \left( 3 d^2 +6 d + 5 - \frac{2 \pi^2}{3} + g_2 \right) (\alpha_s \beta_0)^2 
+ O(\alpha_s^3) \bigg], 
\end{equation}
where $g_2$ is defined by the integral 
\begin{eqnarray}
g_2 &=& 
- \int_0^1 \frac{dx}{x} (1-x) 
\int_0^1 \frac{dy}{y} (1-y) \theta(\sqrt{x}+\sqrt{y}-1) 
\nonumber \\ && 
+ \int_0^1 \frac{dx}{x} \int_0^1 \frac{dy}{y} 
[f(x,y)-(1-x) (1-y) ]  \theta(1-\sqrt{x}-\sqrt{y}). 
\end{eqnarray}
We evaluate this integral numerically to find $g_2 = -3.22467022(9)$. 
It can be seen that the bubble-chain resummation reproduces the fixed-order 
perturbation series in the large $n_f$ limit by comparing the coefficients of 
$(\alpha_s n_f)^n$ for $n=1$ and $2$ in Eqs.~(\ref{eq:R-pert-decoupled}) and
(\ref{eq:R-subtract}).
Equation~(\ref{eq:R-subtract}) also reproduces the leading logarithmic
contributions in Eq.~(\ref{eq:R-pert-decoupled}) in the form 
$(\alpha_s \beta_0 \log \frac{\mu^2}{4 m^2})^n$ for $n=1$ and 2. 
In Eq.~(\ref{eq:R-pert-decoupled}), there is an order-by-order cancellation of the 
renormalization-scale dependence from the two-loop running of $\alpha_s$ with 
$n_f$ active quark flavors and the explicit logarithms of $\mu$. 
On the other hand, in Eq.~(\ref{eq:R-subtract}), the cancellation only occurs
between the one-loop running of $\alpha_s$ and the leading logarithms
$(\alpha_s \beta_0 \log \frac{\mu^2}{4 m^2})^n$ for $n=1$ and 2. Hence, 
Eq.~(\ref{eq:R-subtract}) reproduces 
the subleading logarithm $\alpha_s^2 \log \frac{\mu^2}{4 m^2}$ 
in Eq.~(\ref{eq:R-pert-decoupled}) only in the large $n_f$ limit. 

Our combined result for the ratio $R$, where the 
resummed result and the fixed-order calculation up to NNLO in $\alpha_s$ are
combined, is 
\begin{equation}
\label{eq:R-combined}
R^{\rm Resum+\Delta NNLO}=R^{\rm Resum} + R^{\rm Pert} - \delta R^{\rm Resum}, 
\end{equation}
where $R^{\rm Resum}$, $R^{\rm Pert}$, and $\delta R^{\rm Resum}$ are given in
Eqs.~(\ref{eq:R-resum}), (\ref{eq:R-pert}), and (\ref{eq:R-subtract}),
respectively. 
For $\mu<m$, we use Eq.~(\ref{eq:R-pert-decoupled}) instead of
Eq.~(\ref{eq:R-pert}) to compute $R^{\rm Pert}$.
We also define $R^{\rm Resum+\Delta NLO}$, which is the same as 
$R^{\rm Resum+\Delta NNLO}$, except in $R^{\rm Resum+\Delta NLO}$, 
$R^{\rm Pert}$ and $\delta R^{\rm Resum}$ are computed to NLO in $\alpha_s$. 

Now we discuss the improvement of the perturbative convergence of the combined
result $R^{\rm Resum+\Delta NNLO}$ compared to the fixed-order
calculation $R^{\rm Pert}$. While $R^{\rm Resum}$ 
is valid to all orders in $\alpha_s$ in the large $n_f$ limit, 
$R^{\rm Resum+\Delta NNLO}$ receives radiative corrections from 
$R^{\rm Pert}-\delta R^{\rm Resum}$. 
Then, the perturbative convergence of $R^{\rm Resum+\Delta NNLO}$ is closely 
related to the agreement between $R^{\rm Pert}$ and $\delta R^{\rm Resum}$. 
If we set $\mu = m$, we obtain for the fixed-order perturbative
calculation for $\eta_c$,
\begin{equation}
R^{\rm Pert}_{\eta_c} = R_0 
\bigg[
1 + (9.50-0.427 n_f) \frac{\alpha_s}{\pi} + 
(32.9-6.73 n_f-0.0802 n_f^2) \left(\frac{\alpha_s}{\pi}\right)^2
+O(\alpha_s^3)
\bigg].
\end{equation}
This expression is valid for an arbitrary number of light quark flavors $n_f$, 
except that we only consider
three light quark flavors for the contribution to ``light by light''
contribution to the two-photon decay rate at NNLO in $\alpha_s$ that occurs 
through $Q \bar Q \to gg \to \gamma \gamma$ via a light quark loop, 
which is proportional 
to the sum of squares of light quark fractional charges~\cite{Feng:2015uha}. 
For the decay of $\eta_b$, where we consider four light quark flavors in the 
light by light contribution, we obtain 
\begin{equation}
R^{\rm Pert}_{\eta_b} = R_0 
\bigg[
1 + (9.50-0.427 n_f) \frac{\alpha_s}{\pi} + 
(24.6-6.73 n_f-0.0802 n_f^2) \left(\frac{\alpha_s}{\pi}\right)^2
+O(\alpha_s^3)
\bigg].
\end{equation}
On the other hand, $\delta R^{\rm Resum}$ gives, for both $\eta_c$ and
$\eta_b$,
\begin{subequations}
\begin{eqnarray}
\delta R^{\rm Resum}_{\rm NNA} &=& R_0 
\bigg[
1 + (7.04-0.427 n_f) \frac{\alpha_s}{\pi} + 
(-21.8+2.65 n_f-0.0802 n_f^2) \left(\frac{\alpha_s}{\pi}\right)^2
+O(\alpha_s^3)
\bigg],
\nonumber \\
\\
\delta R^{\rm Resum}_{\rm BFG} &=& R_0 
\bigg[
1 + (9.04-0.427 n_f) \frac{\alpha_s}{\pi} + 
(2.30+1.37 n_f-0.0802 n_f^2) \left(\frac{\alpha_s}{\pi}\right)^2
+O(\alpha_s^3)
\bigg].
\nonumber \\
\end{eqnarray}
\end{subequations}
Here, we chose the gauge-fixing parameter in the BFG method to be $\xi=1$,
which corresponds to the Feynman gauge. 
As expected, $\delta R^{\rm Resum}$ reproduces $R^{\rm Pert}$ only in the large
$n_f$ limit. While it is not at all surprising that $\delta R^{\rm Resum}$ does
not reproduce $R^{\rm Pert}$ beyond the large $n_f$ limit, the size of the
coefficients of order $\alpha_s^2 n_f$ and $\alpha_s^2 n_f^0$ are quite large 
in $R^{\rm Pert}$. If the large $n_f$ limit does not provide a good 
approximation to the fixed-order calculation, the perturbative convergence of 
$R^{\rm Pert}-\delta R^{\rm Resum}$ can be spoiled. 
To inspect the agreement between $R^{\rm Pert}$ and $\delta R^{\rm Resum}$
explicitly, we consider the perturbation series of $R^{\rm Pert}$ and 
$\delta R^{\rm Resum}$ with $n_f = 3$ and $4$ light quark flavors for the 
case of $\eta_c$ and $\eta_b$, respectively. 
For the decay of $\eta_c$ with $n_f = 3$, we obtain
\begin{subequations}
\label{eq:pertcorr_charm}
\begin{eqnarray}
R^{\rm Pert}_{\rm \eta_c} &=& R_0 
\bigg[
1 + 8.22 \frac{\alpha_s}{\pi} + 
12.0 \left(\frac{\alpha_s}{\pi}\right)^2
+O(\alpha_s^3)
\bigg],
\\
\delta R^{\rm Resum}_{\rm \eta_c, NNA} &=& R_0 
\bigg[
1 + 5.76 \frac{\alpha_s}{\pi} - 14.6 \left(\frac{\alpha_s}{\pi}\right)^2
+O(\alpha_s^3)
\bigg],
\\
\delta R^{\rm Resum}_{\rm \eta_c, BFG} &=& R_0 
\bigg[
1 + 7.76 \frac{\alpha_s}{\pi} + 
5.67 \left(\frac{\alpha_s}{\pi}\right)^2
+O(\alpha_s^3)
\bigg].
\end{eqnarray}
\end{subequations}
For the decay of $\eta_b$ with $n_f=4$, we obtain
\begin{subequations}
\label{eq:pertcorr_bottom}
\begin{eqnarray}
R^{\rm Pert}_{\rm \eta_b} &=& R_0 
\bigg[
1 + 7.80 \frac{\alpha_s}{\pi} - 3.61 \left(\frac{\alpha_s}{\pi}\right)^2
+O(\alpha_s^3)
\bigg],
\\
\delta R^{\rm Resum}_{\rm \eta_b, NNA} &=& R_0 
\bigg[
1 + 5.33 \frac{\alpha_s}{\pi} - 12.5 \left(\frac{\alpha_s}{\pi}\right)^2
+O(\alpha_s^3)
\bigg],
\\
\delta R^{\rm Resum}_{\rm \eta_b, BFG} &=& R_0 
\bigg[
1 + 7.33 \frac{\alpha_s}{\pi} + 6.48 \left(\frac{\alpha_s}{\pi}\right)^2
+O(\alpha_s^3)
\bigg].
\end{eqnarray}
\end{subequations}
In all cases, agreement between $R^{\rm Pert}$ and $\delta R^{\rm Resum}$ is
poor at NNLO in $\alpha_s$, even though the difference vanishes in
the large $n_f$ limit. 
Hence, perturbative corrections may not still be in control because of the large
radiative corrections in $R^{\rm Resum+\Delta NNLO}$ 
beyond the large $n_f$ limit.

\section{Numerical Results}
\label{sec:numerical}

We now discuss our numerical results, based on our expression of $R$ in 
Eq.~(\ref{eq:R-combined}). We first describe our numerical inputs. 
We take the heavy-quark mass $m$ to be 1.5~GeV for the charm quark, and 
4.6~GeV for the bottom quark. These values are numerically close to the
one-loop pole mass.
We have a freedom in choosing the values of the heavy-quark mass due to the
ambiguity in the pole mass that is of the order of $\Lambda_{\rm QCD}$. 
The quantity $R^{\rm Resum}$ depends on the heavy-quark mass only through 
$\log \frac{\mu^2}{4 m^2}$. Hence, the dependence on the choice of the
value of the heavy-quark mass $m$ is beyond our accuracy. 
We take the number of light quark flavors to be $n_f = 3$ for $\eta_c$ 
and $n_f = 4$ for $\eta_b$.
In evaluating $R^{\rm Resum}$, we consider both the NNA and the BFG method.
The NRQCD cutoff $\Lambda$ must be chosen between $m v$ and $m$, where 
$v^2 \sim 0.3$ for $\eta_c$ and $v^2 \sim 0.1$ for $\eta_b$. 
Accordingly, we set the central 
value for the NRQCD cutoff to be 1~GeV for $\eta_c$, and 2~GeV for $\eta_b$. 
We take the central value for $\mu$ to be the heavy-quark mass. We compute
$\alpha_s$ in the $\overline {\rm MS}$ renormalization scheme using the {\tt
Mathematica} package {\tt RunDec}~\cite{Chetyrkin:2000yt}. We use $\alpha =
1/137.036$. In the BFG method, we set $\xi=1$, which
corresponds to the Feynman gauge. The choice $\xi=1$ also minimizes the
size of the fixed-order corrections $R^{\rm Pert} - \delta R^{\rm
Resum}$~[see Eqs.~(\ref{eq:pertcorr_charm}, 
\ref{eq:pertcorr_bottom})]. 
In evaluating $R^{\rm Pert}$, we use the expression in Eq.~(\ref{eq:R-pert}) 
for $\mu \ge m$ and use the expression in Eq.~(\ref{eq:R-pert-decoupled})
for $\mu < m$. 

We list the sources of uncertainties that we consider. 
We vary $\Lambda$ by $\pm 25$\% of its central value. 
We vary $\mu$ between 1~and 2~GeV for $\eta_c$, and between 2~and 6~GeV for
$\eta_b$. Because our expression for $R$ in Eq.~(\ref{eq:R-combined}) only 
depends 
on the heavy-quark mass through $\log \frac{\mu^2}{4 m^2}$, where $\mu$ is the
renormalization scale in the $\overline {\rm MS}$ scheme, the dependence of $R$
on $m$ is very mild. Also, the change in $R$ from varying $m$ is equivalent 
to the change in $R$ from varying $\mu$. 
Since we already take into account the uncertainty from the dependence on
$\mu$, we ignore the uncertainty from the dependence on $m$.  
Finally, we estimate the color-octet LDME by using the perturbative estimate
given in Ref.~\cite{Bodwin:2001pt},
\begin{equation}
\frac{\langle \eta_Q | {\cal O}_8 ({}^3S_1) | \eta_Q \rangle}
{\langle \eta_Q | {\cal O}_1 ({}^1S_0) | \eta_Q \rangle}
\sim \frac{v^3 C_F}{\pi N_c}, 
\end{equation}
where we choose 
$v^2 = 0.3$ for $\eta_c$, and $v^2 = 0.1$ for $\eta_b$. The
uncertainty from ignoring the color-octet contribution is then estimated by 
$\pm |R_8|$, where $R_8$ is given by Eq.~(\ref{eq:R8}).
Note that the color-octet matrix element $\langle \eta_Q | 
{\cal O}_8 ({}^3S_1) | \eta_Q \rangle$ depends on $\Lambda$;
hence, there is a correlation between errors from varying $\Lambda$ and ignoring the color-octet matrix element.
We, however, ignore the correlation and add the uncertainties in quadrature.

\subsection{Decay of $\eta_c$}

\begin{figure}
\epsfig{file=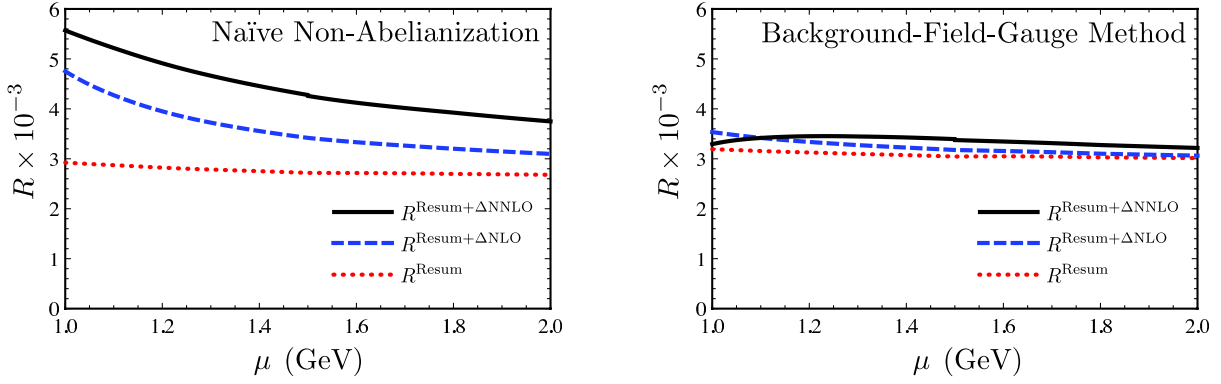,width=16cm}
\caption{
\label{fig:etac-scale}
The renormalization scale dependence of 
$R^{\rm Resum}$ (dotted line), $R^{\rm Resum+\Delta NLO}$ (dashed line) and 
$R^{\rm Resum+\Delta NNLO}$ (black line) 
for $\eta_c$ and $\Lambda = 1$~GeV, 
for the NNA (left) and the BFG method (right). 
}
\end{figure}

We first present our numerical results for the ratio $R$ for $\eta_c$. 
In Fig.~\ref{fig:etac-scale}, we show the dependence on the renormalization
scale $\mu$ of $R^{\rm Resum}$, $R^{\rm Resum+\Delta NLO}$ and 
$R^{\rm Resum+\Delta NNLO}$ at $\Lambda = 1$~GeV. 
For both the NNA and BFG methods, 
$R^{\rm Resum}$ has some dependence on $\mu$ because the renormalization
scale dependence in $R^{\rm Resum}$ only cancels at the one-loop level. 
For the case of NNA, $R^{\rm Resum+\Delta NLO}$ develops a stronger dependence 
on $\mu$ from the fixed-order corrections of relative order $\alpha_s$
in $R^{\rm Pert} - \delta R^{\rm Resum}$. In $R^{\rm Resum+\Delta NNLO}$, the
renormalization scale dependence is slightly worse than 
$R^{\rm Resum+\Delta NLO}$,
because of 
the subleading logarithm of the form 
$\alpha_s^2 \log \frac{\mu^2}{4 m^2}$ in $R^{\rm Pert} - \delta R^{\rm Resum}$. 
For $\mu>m$, there are also uncanceled leading logarithms in 
$R^{\rm Pert} - \delta R^{\rm Resum}$ that are proportional to $n_H$. 
For the case of the BFG method, $R^{\rm Resum}$, $R^{\rm Resum+\Delta NLO}$ and
$R^{\rm Resum+\Delta NNLO}$ all depend on $\mu$ very mildly. This is because in the 
BFG method in the Feynman gauge, there is almost exact cancellation in the fixed-order corrections 
$R^{\rm Pert} - \delta R^{\rm Resum}$, which contain most of the dependence on 
$\mu$. 

\begin{figure}
\epsfig{file=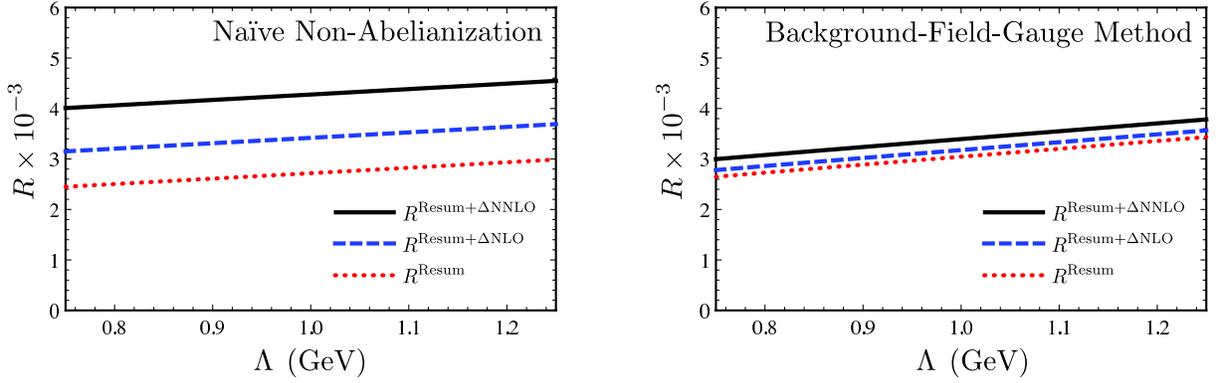,width=16cm}
\caption{
\label{fig:etac-cutoff}
The NRQCD cutoff dependence of 
$R^{\rm Resum}$ (dotted line), $R^{\rm Resum+\Delta NLO}$ (dashed line) and
$R^{\rm Resum+\Delta NNLO}$ (black line)
for $\eta_c$ and $\mu = m$
for the NNA (left) and the BFG method (right). 
}
\end{figure}

In Fig.~\ref{fig:etac-cutoff} we show the dependence on the NRQCD cutoff
$\Lambda$ of $R^{\rm Resum}$, $R^{\rm Resum+\Delta NLO}$ and 
$R^{\rm Resum+\Delta NNLO}$ at
$\mu=m$. In all cases, the dependence on $\Lambda$ is mild, and the
numerical values of $R$ rise slowly with increasing $\Lambda$. As we will see
later, the uncertainty estimated from varying $\Lambda$ is smaller than the
estimated uncertainty from neglecting the color-octet contribution. 

For $\mu=m$, 
the fixed-order corrections in $R^{\rm Pert} - \delta R^{\rm Resum}$
are positive for both the contributions of relative order $\alpha_s$ and 
order $\alpha_s^2$. In NNA, 
$R^{\rm Resum+\Delta NLO}$ is larger than $R^{\rm Resum}$ by about
16\% of the central value of $R^{\rm Resum+\Delta NNLO}$, and 
$R^{\rm Resum+\Delta NNLO}$ is larger than $R^{\rm Resum+\Delta NLO}$ by about
20\% of the central value of $R^{\rm Resum+\Delta NNLO}$. 
In the BFG method, 
$R^{\rm Resum+\Delta NLO}$ is larger than $R^{\rm Resum}$ by about
4\% of the central value of $R^{\rm Resum+\Delta NNLO}$, and
$R^{\rm Resum+\Delta NNLO}$ is larger than $R^{\rm Resum+\Delta NLO}$ by about
6\% of the central value of $R^{\rm Resum+\Delta NNLO}$. 
While the effects of the fixed-order corrections appear less dramatic 
than the effects of the radiative corrections in the fixed-order 
calculation in Ref.~\cite{Feng:2017hlu}, 
the fact that the corrections are larger at NNLO than at NLO in $\alpha_s$ 
implies that the perturbative corrections
may still not be under control. As discussed in the previous section, this is 
related to the large perturbative corrections in the fixed-order corrections
that go beyond the large $n_f$ limit; because the treatment of the renormalon
ambiguities in this work is only valid in the large $n_f$ limit, we have little
or no control over the convergence of the perturbation series beyond the large
$n_f$ limit.

We estimate the uncertainties by varying $\mu$ between 1~and 2~GeV, 
and by varying $\Lambda$ between 0.75 and 1.25~GeV. We also include the
uncertainty for ignoring the color-octet contribution. 
For NNA, we obtain 
\begin{equation}
\label{eq:Retac-NNA}
R^{\rm Resum+\Delta NNLO}_{\eta_c} ({\rm NNA}) 
=  (4.28{}^{+1.29}_{-0.53} \pm 0.27 \pm 0.41) \times 10^3
=  (4.28{}^{+1.38}_{-0.72}) \times 10^3, 
\end{equation}
where the first uncertainty is from $\mu$, the second from $\Lambda$, and the
third uncertainty is from the neglected color-octet contribution.
For BFG, we obtain
\begin{equation}
\label{eq:Retac-BFG}
R^{\rm Resum+\Delta NNLO}_{\eta_c} ({\rm BFG}) 
= ( 3.39{}^{+0.06}_{-0.18} {}^{+0.39}_{-0.40} \pm 0.47) \times 10^{3} 
= ( 3.39{}^{+0.61}_{-0.64}) \times 10^{3}, 
\end{equation}
where the uncertainties are as in NNA. 
Our numerical results for the NNA and the BFG methods are compatible within 
uncertainties. 
We note that $R^{\rm Resum+\Delta NNLO}$ in the NNA method has a large
uncertainty from its strong renormalization-scale dependence for small $\mu$. 

In estimating the uncertainties in our numerical results we have neglected the
possibility that the convergence of the fixed-order corrections in 
$R^{\rm Pert}- \delta R^{\rm Resum}$ may not be in control. 
We roughly estimate the uncertainty from this nonconvergence 
by comparing our numerical results for
$R$ with the series expansion of ${\rm Br} (\eta_c \to \gamma \gamma) = 
1/R_{\rm \eta_c}^{\rm Resum + \Delta NNLO}$ in powers
of $\alpha_s$ through NNLO accuracy. 
For NNA, we obtain ${\rm Br}_{\rm NNA} (\eta_c \to \gamma \gamma) = 
(5.51 \times 10^3)^{-1}$, and for BFG, we obtain
${\rm Br}_{\rm BFG} (\eta_c \to \gamma \gamma) = (3.44 \times 10^3)^{-1}$. 
These values are in agreement with our numerical results in 
Eqs.~(\ref{eq:Retac-NNA}) and (\ref{eq:Retac-BFG}) within uncertainties. 
Therefore, at the current level of accuracy, the uncertainty from the possible
nonconvergence of the fixed-order corrections in 
$R^{\rm Pert}- \delta R^{\rm Resum}$ may not exceed 
our estimated uncertainties. 

We can compare our numerical results with measurements. 
PDG reports two values for the $\eta_c$ branching ratio to two
photons~\cite{Patrignani:2016xqp}. The first PDG value 
${\rm Br} (\eta_c \to \gamma \gamma) = (1.59\pm 0.13) \times 10^{-4}$ 
is from a constrained fit of partial widths. If we take
the inverse, we obtain 
$R^{\rm exp}_{\rm fit} = (6.29{}^{+0.56}_{-0.48}) \times 10^3$. 
The second PDG value is from an average of measurements, which gives 
${\rm Br} (\eta_c \to \gamma \gamma) = (1.9^{+0.7}_{-0.6}) \times 10^{-4}$. 
Taking the inverse gives $R^{\rm exp}_{\rm average} = (5.3{}^{+2.4}_{-1.4}) 
\times 10^3$. 
The two PDG values are compatible with each other due to the large 
uncertainties in $R^{\rm exp}_{\rm average}$. 
The uncertainty in $R^{\rm exp}_{\rm fit}$ is smaller than the uncertainty in 
$R^{\rm exp}_{\rm average}$ or the uncertainties in our numerical results for 
$R$. Note that
$R^{\rm exp}_{\rm average}$ is compatible with our results for $R$ in
Eqs.~(\ref{eq:Retac-NNA}) and (\ref{eq:Retac-BFG}).  
There is, however, a tension between $R^{\rm
exp}_{\rm fit}$ and our numerical results.
We also note that our calculation of $R$
also applies for the $\eta_c (2S)$ state as well. 
The PDG value for the $\eta_c (2S)$ branching ratio to two photons is 
${\rm Br} (\eta_c (2S) \to \gamma \gamma) = (1.9\pm 1.3) \times 10^{-4}$, which
is compatible with the PDG values for the ${\rm Br} (\eta_c \to \gamma \gamma)$
and our results for $R$ in Eqs.~(\ref{eq:Retac-NNA}) and (\ref{eq:Retac-BFG}).

\subsection{Decay of $\eta_b$}

\begin{figure}
\epsfig{file=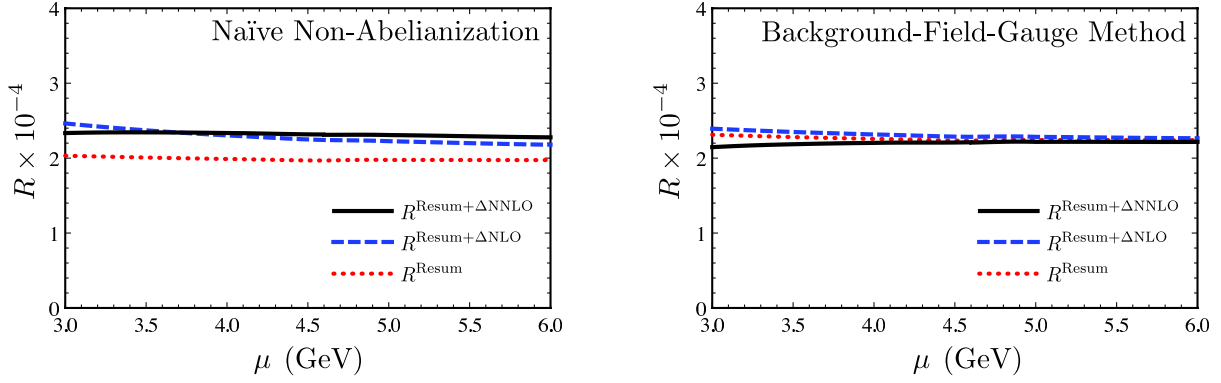,width=16cm}
\caption{
\label{fig:etab-scale}
The renormalization scale dependence of
$R^{\rm Resum}$ (dotted line), $R^{\rm Resum+\Delta NLO}$ (dashed line) and
$R^{\rm Resum+\Delta NNLO}$ (black line)
for $\eta_b$ and $\Lambda = 2$~GeV
for the NNA (left) and the BFG method (right). 
}
\end{figure}

We now present our results for $\eta_b$.
In Fig.~\ref{fig:etab-scale}, we show the dependence on the renormalization
scale $\mu$ of $R^{\rm Resum}$, $R^{\rm Resum+\Delta NLO}$ and $R^{\rm Resum+\Delta NNLO}$ at
$\Lambda = 2$~GeV. Just like for the case of $\eta_c$, 
$R^{\rm Resum}$ has some dependence on $\mu$ because the renormalization
scale dependence in $R^{\rm Resum}$ only cancels at the one-loop level. 
For the case of NNA, $R^{\rm Resum+\Delta NLO}$ develops a stronger dependence on 
$\mu$ from the fixed-order corrections of relative order $\alpha_s$
in $R^{\rm Pert} - \delta R^{\rm Resum}$. This dependence is partially canceled 
by the corrections of relative order $\alpha_s^2$, which contain logarithms of
$\mu$, and $R^{\rm Resum+\Delta NNLO}$ depends on $\mu$ mildly. 
For the case of the BFG method, $R^{\rm Resum}$, $R^{\rm Resum+\Delta NLO}$ and
$R^{\rm Resum+\Delta NNLO}$ all depend on $\mu$ mildly. This is again because the 
choice $\xi=1$ minimizes the size of the fixed-order corrections $R^{\rm Pert} -
\delta R^{\rm Resum}$, which contain most of the dependence on $\mu$. 

\begin{figure}
\epsfig{file=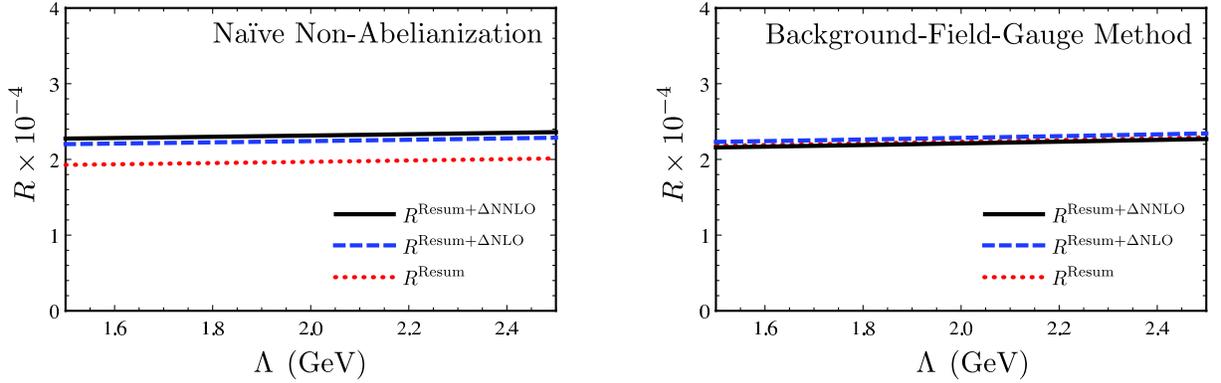,width=16cm}
\caption{
\label{fig:etab-cutoff}
The NRQCD cutoff dependence of
$R^{\rm Resum}$ (dotted line), $R^{\rm Resum+\Delta NLO}$ (dashed line) and
$R^{\rm Resum+\Delta NNLO}$ (black line)
for $\eta_b$ and $\mu = 2$~GeV
for the NNA (left) and the BFG method (right). 
}
\end{figure}

In Fig.~\ref{fig:etab-cutoff} we show the dependence on the NRQCD cutoff
$\Lambda$ of $R^{\rm Resum}$, $R^{\rm Resum+\Delta NLO}$ and $R^{\rm Resum+\Delta NNLO}$ at
$\mu=m$, and $1.5$~GeV~$ \leq \Lambda \leq 2.5$~GeV.
In all cases, the dependence on $\Lambda$ is mild, and the
numerical values of $R$ rise very slowly with increasing $\Lambda$. 
Just like for the case of $\eta_c$, the uncertainty estimated from varying
$\Lambda$ is smaller than the estimated uncertainty from neglecting 
the color-octet contribution. 

For NNA at $\mu=m$ and $\Lambda=2$~GeV, 
the fixed-order corrections in $R^{\rm Pert} - \delta R^{\rm Resum}$
are positive for both the contributions of relative order-$\alpha_s$ and 
order-$\alpha_s^2$. In NNA, 
$R^{\rm Resum+\Delta NLO}$ is larger than $R^{\rm Resum}$ by about
11\% of the central value of $R^{\rm Resum+\Delta NNLO}$, and
$R^{\rm Resum+\Delta NNLO}$ is larger than $R^{\rm Resum+\Delta NLO}$ by about
3\% of the central value of $R^{\rm Resum+\Delta NNLO}$. 
In the BFG method, at $\mu=m$ and $\Lambda=2$~GeV, 
$R^{\rm Resum+\Delta NLO}$ is larger than $R^{\rm Resum}$ by about
2\% of the central value of $R^{\rm Resum+\Delta NNLO}$, 
and $R^{\rm Resum+\Delta NNLO}$ is smaller than $R^{\rm Resum+\Delta NLO}$ 
by about
3\% of the central value of $R^{\rm Resum+\Delta NNLO}$. 
The effects of the fixed-order corrections are much less dramatic than
the corrections to the $\eta_c$ decay rate, thanks to the smaller size of 
$\alpha_s$ and larger $n_f$.

We estimate the uncertainties by varying $\mu$ between 3~and 6~GeV, 
and by varying $\Lambda$ between 1.5~and 2.5~GeV. We also include the
uncertainty for ignoring the color-octet contribution. 
For NNA, we obtain 
\begin{equation}
\label{eq:Retab-NNA}
R^{\rm Resum+\Delta NNLO}_{\eta_b} ({\rm NNA}) 
= ( 2.32{}^{+0.02}_{-0.05} \pm 0.04\pm 0.06 ) \times 10^{4}
= ( 2.32{}^{+0.08}_{-0.09} ) \times 10^{4},
\end{equation}
where the first uncertainty is from $\mu$, the second from $\Lambda$, and the
third uncertainty is from the neglected color-octet contribution. 
For BFG, we obtain
\begin{equation}
\label{eq:Retab-BFG}
R^{\rm Resum+\Delta NNLO}_{\eta_b} ({\rm BFG}) 
= (2.41 {}^{+0.00}_{-0.05}{}^{+0.06}_{-0.05}\pm 0.07) \times 10^{4}
= (2.41 {}^{+0.09}_{-0.10}) \times 10^{4},
\end{equation}
where the uncertainties are as in NNA. 
Our numerical results for the NNA and the BFG methods are compatible 
within uncertainties. 

Even though the fixed-order corrections in $R^{\rm Pert}-\delta R^{\rm Resum}$ 
are not as large as the corrections to the $\eta_c$ decay rate, 
our results may still suffer from nonconvergence. 
We again roughly estimate the uncertainty from this possible nonconvergence by 
comparing our numerical results for $R$ with the series expansion of 
${\rm Br} (\eta_b \to \gamma \gamma) = 1/R_{\eta_b}^{\rm Resum+\Delta NNLO}$ 
in powers of $\alpha_s$ through NNLO accuracy. 
For NNA, we obtain ${\rm Br}_{\rm NNA} (\eta_b \to \gamma \gamma) =
(2.34\times 10^4)^{-1}$, 
and for BFG, we obtain 
${\rm Br}_{\rm BFG} (\eta_b \to \gamma \gamma) =
(2.41\times 10^4)^{-1}$. 
These values are in good agreement with our numerical results in
Eqs.~(\ref{eq:Retab-NNA}) and (\ref{eq:Retab-BFG}). 
This may imply that the uncertainty from the possible nonconvergence of the 
fixed-order corrections in $R^{\rm Pert}-\delta R^{\rm Resum}$ is not
significant for the case of $\eta_b$. 

It is not yet possible to compare our results in Eqs.~(\ref{eq:Retab-NNA}) and
(\ref{eq:Retab-BFG}) with measurements because the partial width $\Gamma_{\eta_b
\to \gamma \gamma}$ has not been observed yet. 
In Ref.~\cite{Chung:2010vz}, the 
authors made use of the heavy-quark spin symmetry to extract $\eta_b$ LDMEs
from the $\Upsilon$ LDMEs and made the prediction 
$\Gamma_{\eta_b \to \gamma \gamma} = 0.512
{}^{+0.096}_{-0.094}$~keV. If we multiply this result to our results for $R$, 
we obtain $\Gamma_{\eta_b} = 11.9 {}^{+2.3}_{-2.2}$~MeV for NNA, and 
$\Gamma_{\eta_b} = 12.4{}^{+2.4}_{-2.3}$~MeV for the BFG method. 
Reference~\cite{Kiyo:2010jm} makes use of the ratio of the leptonic decay rate of 
the $\Upsilon$ to the decay rate of $\eta_b$ into two photons in the 
potential NRQCD effective field theory to predict 
$\Gamma_{\eta_b \to \gamma \gamma}$ from the measured value for 
$\Gamma_{\Upsilon \to e^+ e^-}$. 
The prediction in Ref.~\cite{Kiyo:2010jm} is given by 
$\Gamma_{\eta_b \to \gamma \gamma} = 0.54 \pm 0.15$~keV, which is
compatible with the prediction in Ref.~\cite{Chung:2010vz}. 
If we use this prediction, we obtain $\Gamma_{\eta_b} = 12.5 \pm 3.5$~MeV for
NNA and $\Gamma_{\eta_b} = 13.0 \pm 3.7$~MeV for the BFG method. 
These predictions for the $\eta_b$ decay rate are compatible with the 
PDG value for the $\eta_b$ decay width, which is given by 
$\Gamma_{\eta_b} = 10^{+5}_{-4}$~MeV.

\subsection{Comparison with previous results}

We now compare our numerical results with previous results for $R$. 
In Ref.~\cite{Bodwin:2001pt}, the authors also considered resummation of
bubble-chain contributions to $R$ for the decay of $\eta_c$. The results of Ref.~\cite{Bodwin:2001pt} 
are equivalent to $R^{\rm Resum + \Delta NLO}$, except that in 
Ref.~\cite{Bodwin:2001pt}, the SDC 
was computed by imposing a hard IR cutoff which affects both 
the virtuality and the spacial momentum of the gluon in the perturbative QCD
calculation [Eq.~(\ref{eq:gam-qcd})]. 
The authors of Ref.~\cite{Bodwin:2001pt} identified 
the contribution from the momentum region that was neglected in the 
perturbative QCD calculation 
as the contribution from perturbative NRQCD 
which is regulated by a hard UV cutoff on the gluon momentum. 
The hard cutoff that was used in Ref.~\cite{Bodwin:2001pt} is given by 
$k^2 \leq 4 m^2 \delta$ and $\bm{k}^2 \leq m^2 (2 \sqrt{\delta}-\delta)^2$,
 where $k$ is a gluon momentum, and $\delta = 0.1$. 
If we take $m=1.5$~GeV, we obtain $k^2 \leq
0.9$~GeV${}^2$ and $|\bm{k}| \leq 0.8$~GeV. 
Numerically, the hard cutoff imposed on $|\bm{k}|$ is similar to the 
hard UV cutoff $\Lambda$ that we have employed in this paper, 
although in this work, there is no cutoff on the virtuality of the gluon. 
The main advantage of this work compared to Ref.~\cite{Bodwin:2001pt} 
is that in this work, the appearance and the cancellation of renormalon
ambiguities are explicitly shown by employing dimensional regularization to
regulate infrared divergences. In the numerical results, we have retained the
dependence on the hard cutoff $\Lambda$, whereas the authors of
Ref.~\cite{Bodwin:2001pt} only considered a fixed value of the cutoff. 
We also include the fixed-order corrections at NNLO accuracy in $\alpha_s$.
The authors of Ref.~\cite{Bodwin:2001pt} obtained 
$R_{\eta_c} = (3.01 \pm 0.30 \pm 0.34) \times 10^3$ for NNA, and 
$R_{\eta_c} = (3.26 \pm 0.31 \pm 0.47) \times 10^3$ for the BFG method in the
Feynman gauge. 
The result for the BFG method is compatible with our result in 
Eq.~(\ref{eq:Retac-BFG}), while the result for NNA in 
Ref.~\cite{Bodwin:2001pt} is smaller than our result in 
Eq.~(\ref{eq:Retac-NNA}) by about 30\%.
This difference can be understood 
from the large positive correction at NNLO in $\alpha_s$ 
from $R^{\rm Pert}-\delta R^{\rm Resum}$ that we have 
included in this paper. 

The authors of Ref.~\cite{Feng:2017hlu} presented their numerical results for
${\rm Br} (\eta_c \to \gamma \gamma)$, which is equal to $R^{-1}$, based
on their fixed-order calculation of the inclusive decay rate of $\eta_c$ and
the decay rate of $\eta_c$ into two photons in Ref.~\cite{Feng:2015uha} to NNLO
accuracy in $\alpha_s$. 
The result in Ref.~\cite{Feng:2017hlu} is based on the perturbation 
expansion of ${\rm Br} (\eta_c \to \gamma \gamma)= R^{-1}$ to NNLO in 
$\alpha_s^2$. 
By varying the renormalization scale $\mu$ from 1~GeV
to 3 times the charm quark mass, the authors of Ref.~\cite{Feng:2017hlu} 
obtained ${\rm Br} (\eta_c \to \gamma \gamma) = (3.1$---$3.3) \times
10^{-4}$, which gives $R_{\eta_c} = (3.0$---$3.2)\times 10^3$. This result 
is compatible with our result in Eq.~(\ref{eq:Retac-BFG}) in the BFG method,
but is smaller than our result in Eq.~(\ref{eq:Retac-NNA}) in NNA. 
Also, the uncertainty in the result in Ref.~\cite{Feng:2017hlu} is smaller 
than the uncertainties in our results, due to the cancellation of the 
renormalization-scale dependence at the two-loop level in the fixed-order
calculation. Moreover, 
the uncertainty from the color-octet contribution at relative order $v^3$ has 
been neglected in Ref.~\cite{Feng:2017hlu}. 
One can obtain a different numerical result if one considers the 
perturbation series of $R_{\eta_c}=[{\rm Br} (\eta_c \to \gamma \gamma)]^{-1}$, 
which is given by Eq.~(\ref{eq:R-pert}). If we use Eq.~(\ref{eq:R-pert}) 
we obtain $R_{\eta_c} =4.9 \times 10^3$ at $\mu = m$.
This disagrees with the numerical results in 
Ref.~\cite{Feng:2017hlu}, and the discrepancy is much larger than the 
uncertainties estimated in Ref.~\cite{Feng:2017hlu} by varying the 
renormalization scale $\mu$.
The difference between the numerical results based on the perturbation series 
of $R_{\eta_c}$ and the one based on the perturbation series of 
${\rm Br} (\eta_c \to \gamma \gamma)$ shows that the nonconvergence of the 
perturbation series generates a sizable ambiguity. This is consistent with 
our estimate of the leading renormalon uncertainty in 
Eq.~(\ref{eq:leading_ambiguity}). 
Our results in Eqs.~(\ref{eq:Retac-NNA}) and (\ref{eq:Retac-BFG}) also 
suffer from nonconvergence of the fixed order corrections in $R^{\rm Pert} -
\delta R^{\rm Resum}$, because we have no control over the convergence of
the perturbation series beyond the large $n_f$ limit. 
We have roughly estimated the uncertainty from this nonconvergence by comparing
our numerical results with the series expansion of 
${\rm Br} (\eta_c \to \gamma \gamma) = 
1/R_{\eta_c}^{\rm Resum+\Delta NNLO}$ in powers of $\alpha_s$ through 
NNLO accuracy. 
We have found that our rough estimate of the uncertainty from the 
nonconvergence does not exceed the uncertainties in our numerical results. 

In Ref.~\cite{Feng:2017hlu}, the authors also presented their numerical results
based on the perturbative expression of ${\rm Br} (\eta_b \to \gamma \gamma)$
to NNLO accuracy. They obtained ${\rm Br} (\eta_b \to \gamma \gamma) = 
(4.8 \pm 0.7) \times 10^{-5}$. 
If we take the inverse we obtain $R_{\eta_b} = (2.1 {}^{+0.4}_{-0.3}) \times 10^4$, 
which is compatible with our numerical results in 
Eqs.~(\ref{eq:Retab-NNA}) and~(\ref{eq:Retab-BFG}) within uncertainties. 
The uncertainties in Eqs.~(\ref{eq:Retab-NNA}) and~(\ref{eq:Retab-BFG}) are much 
smaller than the result in Ref.~\cite{Feng:2017hlu} because the bubble-chain 
resummation reduces the dependence on the renormalization scale $\mu$ 
compared to the fixed-order calculation. 
If we use
the perturbative expression of $R$ in Eq.~(\ref{eq:R-pert}), which is valid to
NNLO accuracy in $\alpha_s$, we obtain $R_{\eta_b}=2.39\times 10^4$ at $\mu=m$, 
which agrees with the numerical result in Ref.~\cite{Feng:2017hlu} within 
uncertainties. 
The relative discrepancy between the numerical result from the perturbative 
expression of the branching ratio into two photons and the numerical result 
from the perturbative expression of $R$ is smaller in the case of $\eta_b$
compared to the case of $\eta_c$.  
This can be understood from our estimate of the leading renormalon ambiguity 
[Eq.~(\ref{eq:leading_ambiguity})] : since the decay of $\eta_b$
occurs at a higher energy scale than the decay of $\eta_c$, 
the renormalon ambiguity is suppressed compared to the case of $\eta_c$. 
Nevertheless, the ambiguity is still sizable compared to the estimated 
uncertainties in our numerical results in 
Eqs.~(\ref{eq:Retab-NNA}) and~(\ref{eq:Retab-BFG}). 
Therefore,
even for the case of $\eta_b$, resumming large perturbative corrections is
crucial in obtaining a reliable theoretical prediction. 

In Ref.~\cite{Du:2017lmz}, the authors applied the principle of maximal
conformality (PMC), which is a method for choosing the renormalization scale 
$\mu$ for a given perturbation series, to the perturbative expression for $R$ 
to NLO accuracy. The authors of Ref.~\cite{Du:2017lmz} claim that, by applying 
the PMC, the $\beta$ function appearing in the perturbation series, which are
associated with the running of $\alpha_s$, is absorbed into the running 
coupling, and the
convergence of the perturbation series is improved. 
When they include the relative order-$\alpha_s$ and order-$\alpha_s v^2$
corrections to $R$, they obtain, after applying the PMC, 
$R = (6.09{}^{+0.21}_{-0.19}{}^{+0.58}_{-0.52}) 
\times 10^{3}$. This result is very different from our results in
Eqs.~(\ref{eq:Retac-NNA}) and (\ref{eq:Retac-BFG}), 
which include explicitly the leading-logarithmic corrections 
of the form $(\alpha_s \beta_0 \log \frac{\mu^2}{4 m^2})^n$ to all orders in
$\alpha_s$. 
It is worth noting that, unlike the expressions for $R$ at 
LO and NLO accuracies, the perturbative expression for $R$ at NNLO accuracy no 
longer suffers from the severe dependence on the renormalization
scale~\cite{Feng:2017hlu}. Even if we consider a wide range of the 
renormalization scale, as the authors of Ref.~\cite{Feng:2017hlu} have done, 
it is not possible to obtain a value of $R$ that is close to the result of 
Ref.~\cite{Du:2017lmz} if one uses the expression for $R$ at NNLO accuracy. 
Shortcomings of the PMC approach have been discussed in 
Ref.~\cite{Kataev:2014jba}.

\section{Summary and Discussion}
\label{sec:summary}

In this paper we have presented an analysis of the ratio $R$ of the
inclusive decay rate of the $\eta_Q$ meson to the partial decay rate 
into two photons, where $Q =c$ or $b$. 
In the calculation of the short-distance coefficients, 
we resum large perturbative corrections in the form 
$(\alpha_s \beta_0)^n$ to all orders in $\alpha_s$
by including contributions from bubble chain insertions in the gluon
propagator. This bubble-chain resummation reproduces fixed-order perturbative
calculations in the large $n_f$ limit. 
This resummation has been done in Ref.~\cite{Bodwin:2001pt} by imposing an 
infrared cutoff 
in the perturbative calculations. In this work, we regulate the infrared
divergences using dimensional regularization, so that the appearance of the
renormalon ambiguity in the perturbative QCD calculation and the cancellation
of the ambiguity in the factorization formula can be
seen explicitly. We use na\"ive non-Abelianization and the background-field
gauge method to carry out the resummation, which are unambiguous procedures for
resumming bubble chains. 

We confirmed that, by using the factorization formula valid to relative order 
$v^3$, the leading renormalon ambiguity of infrared origin 
that arises from 
the perturbative QCD calculation is reproduced in the perturbative NRQCD
calculation, and therefore, the short-distance coefficients are free of
infrared
renormalon ambiguities. We also showed that, if we use dimensional 
regularization to regulate the ultraviolet divergences in NRQCD,  
the color-octet LDME suffers from renormalon 
ambiguities of ultraviolet origin, but the ambiguity cancels in the 
factorization formula. Since it is not known how to compute the
color-octet LDME reliably, and it is known that the 
color-octet LDME is suppressed by $v^3$ compared to the
leading-order color-singlet LDME, the 
color-octet contribution is often neglected in the factorization formula. 
However, in a resummed calculation, the neglect of the color-octet contribution 
results in a sizable ambiguity in the factorization formula. 
We argued that, for phenomenological applications, we obtain a more useful
factorization formula if we use hard-cutoff regularization to regulate the 
ultraviolet divergences in NRQCD where such ambiguity no longer appears.

Our result for the resummed calculation of $R$ is shown in 
Eq.~(\ref{eq:R-resum}). We combine our result with the calculation in
fixed-order perturbation theory
to next-to-next-to-leading order accuracy in $\alpha_s$ 
[Eq.~(\ref{eq:R-pert})]~\cite{Barbieri:1979be, Hagiwara:1980nv, 
Czarnecki:2001zc, Feng:2015uha, Feng:2017hlu}. 
The expression for the combined result is shown in
Eq.~(\ref{eq:R-combined}). We use the expression in Eq.~(\ref{eq:R-combined})
in our numerical analysis. 

In our numerical analysis, we estimated uncertainties by varying the
renormalization scale and the NRQCD ultraviolet cutoff. 
We also included the effect of the color-octet contribution by estimating the 
size of the uncalculated color-octet LDME. 
Our numerical results for the ratio $R$ for the decay of $\eta_c$ are given in
Eqs.~(\ref{eq:Retac-NNA}) and (\ref{eq:Retac-BFG}), which are computed in the 
na\"ive non-Abelianization and the background-field gauge method in the Feynman
gauge, 
respectively. The results in Eqs.~(\ref{eq:Retac-NNA}) and 
(\ref{eq:Retac-BFG}) agree within uncertainties. 
Our numerical results for $\eta_c$ are 
compatible with the PDG value for ${\rm Br} (\eta_c \to \gamma \gamma)$ 
that was obtained by taking averages of
measurements. However, our results disagree with the PDG value for 
${\rm Br} (\eta_c \to \gamma \gamma)$ that was obtained from constrained fits. 
For the decay of $\eta_b$, our numerical results for the ratio $R$ are given in
Eqs.~(\ref{eq:Retab-NNA}) and (\ref{eq:Retab-BFG}), which are computed in the
na\"ive non-Abelianization and the background-field gauge method in the Feynman
gauge, respectively.
Again, the numerical results in Eqs.~(\ref{eq:Retab-NNA}) and
(\ref{eq:Retab-BFG}) agree within uncertainties. 
Since the decay of $\eta_b$ into two photons is yet to be measured, we cannot
compare our results for $R$ directly with measurements for the case of
$\eta_b$. By using predictions of $\Gamma_{\eta_b \to \gamma \gamma}$ in
Refs.~\cite{Chung:2010vz, Kiyo:2010jm}, we have obtained predictions of 
$\Gamma_{\eta_b}$ that is compatible with the current measurement of the
$\eta_b$ decay rate. 

We have compared our numerical results with previous calculations of $R$ 
in Ref.~\cite{Bodwin:2001pt}, where the authors also considered bubble-chain 
resummation, and the results based on fixed-order perturbation theory in 
Ref.~\cite{Feng:2017hlu}. 
In Ref.~\cite{Bodwin:2001pt}, the authors made predictions of $R$ for the 
decay of $\eta_c$ by combining the resummed result, which was computed by 
using a fixed infrared cutoff, with the fixed-order 
calculation valid to next-to-leading order in $\alpha_s$. 
Our numerical results agree with the results in Ref.~\cite{Bodwin:2001pt}
for the background-field gauge method [Eq.~(\ref{eq:Retac-BFG})], but there 
is tension in the result in na\"ive non-Abelianization 
[Eq.~(\ref{eq:Retac-NNA})]. This discrepancy is mostly from the inclusion of 
the fixed-order corrections $R^{\rm Pert} - \delta R^{\rm Resum}$ 
at next-to-next-to-leading order in $\alpha_s$ 
[Eq.~(\ref{eq:R-combined})].
While the authors of Ref.~\cite{Bodwin:2001pt} included the effect of the 
color-octet contribution in the uncertainties, the 
uncertainty from the dependence on the infrared cutoff was neglected. 
The numerical results for $R$ for $\eta_c$ in Ref.~\cite{Feng:2017hlu} is also
compatible with our results in the background-field gauge method 
[Eq.~(\ref{eq:Retac-BFG})], but disagrees with our results in na\"ive 
non-Abelianization [Eq.~(\ref{eq:Retac-NNA})]. The uncertainties in the 
result of Ref.~\cite{Feng:2017hlu} is smaller than the uncertainties in 
our results because in the fixed-order calculation, the dependence on the 
renormalization scale cancels at two-loop accuracy, and the uncertainty 
from the uncalculated color-octet contribution is neglected. Also, the 
fixed-order calculation in Ref.~\cite{Feng:2017hlu} suffers from a sizable
uncertainty from the nonconverging
 perturbation series, which, for the case of $\eta_c$, 
can be of relative order one. 
The authors of Ref.~\cite{Feng:2017hlu} also made a prediction of $R$ for 
the decay of $\eta_b$, which agrees with our results 
in Eqs.~(\ref{eq:Retab-NNA}) and (\ref{eq:Retab-BFG}) within uncertainties. 
The uncertainties in our results are smaller than the uncertainty in the 
prediction from fixed-order perturbation theory in Ref.~\cite{Feng:2017hlu}. 
Although in the case of $\eta_b$, the estimated renormalon ambiguity in the 
perturbation series for $R$ is smaller than the case of $\eta_c$, our 
estimate of the ambiguity is larger than the uncertainties in our numerical 
results in Eqs.~(\ref{eq:Retab-NNA}) and (\ref{eq:Retab-BFG}). 
Therefore, we conclude that resummation is necessary in order to obtain an 
accurate prediction of $R$ for the decay of $\eta_b$. 

In Ref.~\cite{Du:2017lmz}, the authors applied the principle of maximal 
conformality to the perturbative expression of $R$ for the decay of $\eta_c$ 
valid to next-to-leading order in $\alpha_s$. 
The authors of Ref.~\cite{Du:2017lmz} claimed that
a resummed perturbative expression can be obtained, where the $\beta$ 
function that is associated with the
running of $\alpha_s$ 
are absorbed into the coupling, by using
the principle of maximal
conformality. However, we find that our resummed result disagrees with the 
result in Ref.~\cite{Du:2017lmz}. The result in Ref.~\cite{Du:2017lmz} also 
disagrees with the result from fixed-order perturbation theory valid to 
next-to-next-to-leading order in $\alpha_s$ in Ref.~\cite{Feng:2017hlu}. 

It is noticeable that the uncertainty estimated from the neglected 
color-octet contribution is quite
significant for both $\eta_c$ and $\eta_b$. This suggests that in order to have
a more precise prediction of $R$, it is necessary to include color-octet 
contributions in $R$. Including the color-octet contribution may also reduce 
the uncertainty from the NRQCD cutoff dependence, because the dependence on the
NRQCD cutoff cancels in the factorization formula between the color-singlet and
color-octet contributions. Since currently it is not known how to calculate
the color-octet matrix element reliably, it would be important to develop new
ideas to investigate the nature of the color-octet matrix element in NRQCD and
other effective field theories such as potential NRQCD, which may help 
constrain the color-octet contribution. 

In our numerical results we included corrections from fixed-order
calculations to next-to-next-to-leading order in $\alpha_s$. 
While the bubble-chain resummation 
reproduces the fixed-order corrections in the large $n_f$ limit, 
the fixed-order corrections are still found to be 
significant beyond the large $n_f$ limit; even after the bubble-chain
resummation, the numerical results for $R$ for the decay of $\eta_c$ 
suggest nonconvergence of perturbative corrections to persist beyond the large
$n_f$ limit. Therefore, in order to gain control over the perturbation series 
of $R$ for the decay of $\eta_c$, 
it may be necessary to consider renormalon ambiguities beyond the
large $n_f$ limit.
By inspecting the $n_f$-dependence of the fixed-order corrections to 
the electromagnetic decay rates $\Gamma_{\eta_c \to \gamma \gamma}$ and 
$\Gamma_{J/\psi \to e^+ e^-}$, which are available up to 
two~\cite{Barbieri:1979be, Czarnecki:2001zc, Feng:2015uha} and three
loops~\cite{Barbieri:1975ki, Celmaster:1978yz, Beneke:1997jm, Czarnecki:1997vz,
Marquard:2014pea}, respectively, we find that 
the fixed-order corrections are also significant beyond
the large $n_f$ limit in those electromagnetic decay rates. 
Hence, the bubble-chain resummation calculations of those decay rates 
in Ref.~\cite{Braaten:1998au} seem to fail to reproduce the fixed-order 
calculations.

We have examined a method that is often employed for computing renormalon 
singularities in the heavy-quark pole mass described in 
Refs.~\cite{Pineda:2001zq, Komijani:2017vep}, 
where the renormalon ambiguities that scale like powers of 
$\Lambda_{\rm QCD}$ 
are subtracted from the divergent perturbation series.
This method has an advantage that it does not rely on the large-$n_f$
limit.
We have found that a na\"ive application of the method in 
Refs.~\cite{Pineda:2001zq, Komijani:2017vep} to the electromagnetic decay rates 
$\Gamma_{\eta_c \to \gamma \gamma}$, $\Gamma_{J/\psi \to e^+ e^-}$ and 
the inclusive decay rate of $\eta_c$ lead to estimates of the perturbative
series that are in poor agreement with the fixed-order corrections. 

By combining the resummed result for $R$ and the fixed-order calculations 
valid up to next-to-next-to-leading order in $\alpha_s$, we have obtained 
precise predictions of $R$ for the decay of $\eta_b$ with uncertainties 
that could be as small as 5\%. 
Therefore, the measurement of $\Gamma_{\eta_b\to\gamma \gamma}$ 
in ongoing and future experiments is highly anticipated. 
We also look forward to improved experimental measurements for 
the decay rate of $\eta_b$, as well as the total and partial decay rates of 
$\eta_c$.

\appendix

\section{Computation of $\bm{T(t,\tau)}$}
\label{Appendix:T-x-y}

In this appendix, we calculate $T(t,\tau)$ defined in Eq.~(\ref{eq:T_t-tau:def}),
\begin{equation}
  T(t,\tau) = \frac{1}{\pi^2}\int_0^1\int_0^1 dx\,  dy\, {\rm Im}\left[\frac{x^{-t}\,e^{i\pi t}}{x+i\varepsilon}\right]
  {\rm Im}\left[\frac{y^{-\tau}\,e^{i\pi \tau}}{y+i\varepsilon}\right]f(x,y)\, \theta(1-\sqrt{x}-\sqrt{y})\nonumber ,
\end{equation}
with $f(x,y)$ given in Eq.~(\ref{eq:f_x-y:def}).
We need to calculate the derivatives of $T(t,\tau)$ at $t=\tau=0$. 
As infrared regulators we assume $\tau<0$ and $t<0$ so that the integral
over $x$ and $y$ become finite. We then drop the small $i\varepsilon$
terms, and we write $T$ as
\begin{equation}
  T(t,\tau) = \frac{1}{\pi^2} \sin(\pi t)\sin(\pi \tau)
  \int_0^1 \int_0^1 dx\,dy\, x^{-t-1}\,y^{-\tau-1}\,f(x,y)\,\theta(1-\sqrt{x}-\sqrt{y})\, .
  \label{eq:T-t-tau:1}
\end{equation}
Using change of variables $y = (1-\sqrt{x})^2 z$, we obtain
\begin{align}
  T(t,\tau) &= \frac{1}{\pi^2} \sin(\pi t)\sin(\pi \tau)
  \int_0^1 \int_0^1 dx\,dz\,x^{-t-1}\,z^{-\tau-1}\, (1-\sqrt{x})^{-2\tau}\,  f(x,(1-\sqrt{x})^2 z)\, .
  \label{eq:T_t-tau:2}
\end{align}
Let us now focus on $f(x,y)$, which can be written as 
\begin{align}
  f(x,y)&=\frac{\Bigl[\big((1-\sqrt{x})^2-y\big)\,\big((1+\sqrt{x})^2-y\big)\Bigr]^{3/2}}{(1-x-y)^2}\nonumber \\
	&=(1-x)\,(1-z)^{3/2}\,\frac{(1-\xi^2 z)^{3/2}}{(1-\xi z)^2}\nonumber \\
	&=(1-x)\,(1-z)^{3/2}\,\sum_{j,k=0}^{\infty} C_{jk}\, \xi^{2k+j}\,z^{k+j}\, ,
  \label{eq:f_x-y:1}
\end{align}
where 
\begin{align}
  \xi &= \frac{1-\sqrt{x}}{1+\sqrt{x}}\,,\\
  C_{jk} &= (j+1)\,\frac{\Gamma(k-3/2)}{\Gamma(-3/2)\Gamma(k+1)}\,  \, .
\end{align}
Plugging Eq.~(\ref{eq:f_x-y:1}) in Eq.~(\ref{eq:T_t-tau:2}) we obtain 
\begin{align}
  T(t,\tau) &= \frac{1}{\pi^2}\,\sin(\pi t) \sin(\pi \tau) \sum_{j,k=0}^{\infty} C_{jk}
  \int_0^1 dz\,z^{k+j-\tau-1} (1-z)^{3/2}
  \int_0^1 dx\, x^{-t-1}\,(1-x)\, (1-\sqrt{x})^{-2\tau}\,\xi^{2k+j} \nonumber \\
  &= \frac{1}{\pi^2}\,\sin(\pi t) \sin(\pi \tau) \sum_{j,k=0}^{\infty} C_{jk}\, B(\frac{5}{2},\, k{+}j{-}\tau)
  \int_0^1 dx\, x^{-t-1}\,(1-x)\, (1-\sqrt{x})^{-2\tau}\,\xi^{2k+j}\nonumber \\
  &= \frac{2}{\pi^2}\,\sin(\pi t) \sin(\pi \tau) \sum_{j,k=0}^{\infty} C_{jk}\, B(\frac{5}{2},\,k{+}j{-}\tau)\, 
  \int_0^1 dX\, X^{-2t-1}\, (1-X)^{2k+j+1-2\tau}\,(1+X)^{1-2k-j} \nonumber \\
  &= \frac{2}{\pi^2}\,\sin(\pi t) \sin(\pi \tau) \sum_{j,k=0}^{\infty} C_{jk}\, B(\frac{5}{2},\,k{+}j{-}\tau)\nonumber \\
  &\quad \times B(-2t,\, 2k{+}j{+}2{-}2\tau)\,  F(2k{+}j{-}1,\, -2t;\, 2k{+}j{+}2{-}2\tau{-}2t;\, -1),
  \label{eq:T_t-tau:3}
\end{align}
with the hypergeometric function
\begin{equation}
  F(\alpha,\beta; \gamma; z) = 
  \frac{1}{B(\beta,\gamma-\beta)}\int_0^1 dx\,x^{\beta-1} (1-x)^{\gamma-\beta-1} (1 - z x)^{-\alpha} .
\end{equation}
Using the identities 
\begin{align}
  F(\alpha, \beta; \gamma; z ) &= F(\beta, \alpha; \gamma; z) ,\\
  F(\alpha, \beta; \gamma; -1) &= 2^{-\alpha}\, F(\alpha, \gamma-\beta; \gamma; \frac{1}{2}) ,
  \label{eq:hyp2f1:identity}
\end{align}
we replace 
\begin{equation}
  F(2k{+}j{-}1,\, -2t;\, 2k{+}j{+}2{-}2\tau{-}2t;\, -1)\, \to\,
  2^{2t}\, F(-2t,\, 3{-}2t{-}2\tau;\, 2k{+}j{+}2{-}2\tau{-}2t;\, \frac{1}{2}) ;\nonumber
\end{equation}
Eq.~(\ref{eq:T_t-tau:3}) reads
\begin{align} 
  T(t,\tau) 
  &= \frac{2\,}{\pi^2}\, \sin(\pi t)\, \sin(\pi \tau)\, \Gamma(-2t)\,2^{2t} 
  \sum_{j,k=0}^{\infty} (j+1)\,\frac{\Gamma(k{-}3/2)}{\Gamma(-3/2)\,\Gamma(k{+}1)}\,
   \frac{\Gamma(5/2)\,\Gamma(k{+}j{-}\tau)}{\Gamma(5/2{+}k{+}j{-}\tau)}\nonumber \\
  &\quad \times\, \frac{\Gamma(2k{+}j{+}2{-}2\tau)}{\Gamma(2k{+}j{+}2{-}2\tau{-}2t)}\,
  F(-2t,\, 3{-}2t{-}2\tau;\, 2k{+}j{+}2{-}2\tau{-}2t;\, \frac{1}{2})
  \label{eq:T_t-tau:4}\, .
\end{align}

Using Legendre's duplication formula and Euler's reflection formula, 
\begin{align}
  \Gamma(-2t) &= \frac{2^{-2t}\,\Gamma(-t)\,\Gamma(1/2-t)}{2\,\Gamma(1/2)}\, ,
  \label{eq:Legendre-duplication} \\
  \sin(\pi t) &= \frac{-\pi}{\Gamma(1+t)\,\Gamma(-t)}\, ,
  \label{eq:Euler-reflection}
\end{align}
we now write Eq.~(\ref{eq:T_t-tau:4}) as 
\begin{align}
  T(t,\tau) &= 
  \frac{1}{\Gamma(1{+}t)\,\Gamma(1{+}\tau)}\,\frac{\Gamma(1/2{-}t)}{\Gamma(1/2)\,\Gamma(-\tau)}\, 
  \sum_{j,k=0}^{\infty} (j+1)\,\frac{\Gamma(k{-}3/2)}{\Gamma(-3/2)\,\Gamma(k{+}1)}\,
   \frac{\Gamma(5/2)\,\Gamma(k{+}j{-}\tau)}{\Gamma(5/2{+}k{+}j{-}\tau)}\nonumber \\
  &\quad \times\, \frac{\Gamma(2k{+}j{+}2{-}2\tau)}{\Gamma(2k{+}j{+}2{-}2\tau{-}2t)}\,
  F(-2t,\, 3{-}2t{-}2\tau;\, 2k{+}j{+}2{-}2\tau{-}2t;\, \frac{1}{2})
  \label{eq:T_t-tau:5}\, .
\end{align}
Note that we have  
\begin{equation}
  T(t,0) = T(0,t) 
  = \frac{\sin(\pi t)}{\pi\, t\, (1-t)}\, .  
  \label{eq:T:at:t=0}
\end{equation}
This relation immediately yields $T(0,0) = 1$. 

The expression given in Eq.~(\ref{eq:T_t-tau:5}) contains summations over $j$ and $k$.
Now we discuss how to improve the radius of convergences of the sums by adding and subtracting some terms that can be summed up analytically.
First, instead of using Eq.~(\ref{eq:f_x-y:1}), we expand $f$ as
\begin{align}
  f(x,y) &= (1-x)\,(1-z)^{3/2}\Bigg\{\sum_{j,k=0}^{\infty} C_{jk}
   \Big[ z^{k+j}\xi^{2k+j} - f_0 - f_1\,(z \xi)^{k+j} \Big] \nonumber \\
   &\qquad\qquad + f_0\, (1-z)^{-1/2} 
   + f_1\sum_{l=0}^{\infty} \frac{\Gamma(l+\frac{1}{2})}{\Gamma(\frac{1}{2})\,\Gamma(l+1)} (z \xi)^{l}
   \Bigg\} ,
  \label{eq:f_x-y:2}
\end{align}
where $f_0$ and $f_1$ can be any constants or any functions of $t$ and $\tau$ that are analytic
in the vicinity of the origin of the complex-$t$ and complex-$\tau$ planes.
Then, we write Eq.~(\ref{eq:T_t-tau:5}) as 
\begin{equation}
  T(t,\tau) = T_0(t,\tau) + T_1(t,\tau) + T_2(t,\tau)
  \label{eq:T-t-tau:6}\, ,
\end{equation}
where 
\begin{align}
  T_0(t,\tau) &\equiv f_0\,
  \frac{-\tau}{\Gamma(1+t)\,\Gamma(1+\tau)}\,
  \frac{\Gamma(\frac{7}{2})}{\Gamma(\frac{3}{2})}\,
  \frac{\Gamma(\frac{3}{2}-t)}{\Gamma(\frac{7}{2}-2 t)}
  \frac{\Gamma(2 -2 t)}{\Gamma(3 -\tau - 2 t)}
  \\
  T_1(t,\tau) &\equiv 
  \frac{1}{\Gamma(1+t)\,\Gamma(1+\tau)}\,\frac{\Gamma(\frac{1}{2}-t)}{\Gamma(\frac{1}{2})\,\Gamma(-\tau)}\, 
  \sum_{l=0}^{\infty} \frac{\Gamma(l+\frac{1}{2})}{\Gamma(\frac{1}{2})\,\Gamma(l+1)}\,
   \frac{\Gamma(\frac{5}{2})\,\Gamma(l-\tau)}{\Gamma(\frac{5}{2}+l-\tau)}\nonumber \\
  &\quad \times\,\Bigg\{ f_1\,\frac{\Gamma(l+2-2\tau)}{\Gamma(l+2-2\tau-2t)}\, F(-2t,\, 3 -2t -2\tau;\, l+2-2\tau-2t;\, \frac{1}{2})
  \nonumber \\
   &\qquad\quad - f_0\,\frac{(l-\tau)\,\Gamma(\frac{5}{2}+l-\tau)}{\Gamma(\frac{7}{2}+l-\tau-2t)} \Bigg\}
  \\ 
  T_2(t,\tau) &\equiv
  \frac{1}{\Gamma(1+t)\,\Gamma(1+\tau)}\,\frac{\Gamma(\frac{1}{2}-t)}{\Gamma(\frac{1}{2})\,\Gamma(-\tau)}\, 
  \sum_{j,k=0}^{\infty} (j+1)\,\frac{\Gamma(k-\frac{3}{2})}{\Gamma(-\frac{3}{2})\,\Gamma(k+1)}\,
   \frac{\Gamma(\frac{5}{2})\,\Gamma(k+j-\tau)}{\Gamma(\frac{5}{2}+k+j-\tau)}\nonumber \\
  &\quad \times\,\Bigg\{ \frac{\Gamma(2k+j+2-2\tau)}{\Gamma(2k+j+2-2\tau-2t)}\, F(-2t,\, 3 -2t -2\tau;\, 2k+j+2-2\tau-2t;\, \frac{1}{2}) 
  \nonumber \\
   &\qquad\quad - f_1\,\frac{\Gamma(k+j+2-2\tau)}{\Gamma(k+j+2-2\tau-2t)}\, F(-2t,\, 3 -2t -2\tau;\, k+j+2-2\tau-2t;\, \frac{1}{2})
  \nonumber \\
   &\qquad\quad - f_0\, \frac{2t k\,\Gamma(\frac{5}{2}+k+j-\tau)}{\Gamma(\frac{7}{2}+k+j-\tau-2t)} \Bigg\}
  \label{eq:S-t-tau_def}\, .
\end{align}
Setting $f_0$ and $f_1$ to unity, the above expression for $T(t,\tau)$ is convergent for all $t<1$.
In particular, one can verify that $T(t,\tau)$ does not have any singularity at $t=\frac{1}{2}$
although it contains a factor of $\Gamma(t-1/2)$.
One can also show that
\begin{equation}
  \lim\limits_{\tau\to1} (1-\tau) T(t,\tau) = -\frac{3}{\pi}\,\sin(\pi t)\, , 
  \label{eq:T-residue-at-tau=1}
\end{equation}
and by symmetry argument 
\begin{equation}
  \lim\limits_{t\to1} (1-t) T(t,\tau) = -\frac{3}{\pi}\,\sin(\pi \tau)\, . 
  \label{eq:T-residue-at-t=1}
\end{equation}


\begin{acknowledgments}

We thank Antonio Vairo for his collaboration on this project,
his valuable comments and careful reading of the manuscript.
The work of N.~B.\ is supported by 
the DFG and the NSFC through funds provided to the
Sino-German CRC 110 ``Symmetries and the Emergence of Structure in QCD'' (NSFC
Grant No.
11621131001). 
N.~B.\ also acknowledges support from the DFG cluster of excellence
``Origin and structure of the universe'' (www.universe-cluster.de).
This work was supported in part by the German Excellence Initiative and the European Union Seventh Framework Program under Grant
Agreement No.~291763 as well as the European Union's Marie Curie COFUND program (J.K.).
The work of H.~S.~C.\ is supported by the Alexander von Humboldt Foundation. 
The work of H.~S.~C.\ is also supported by the DFG cluster of excellence 
``Origin and Structure of the Universe'' (www.universe-cluster.de).

\end{acknowledgments}


\end{document}